\pdfoutput=1

\documentclass[tog]{acmsiggraph}
\usepackage{amsmath}
\usepackage{amsfonts}
\usepackage{amssymb}
\usepackage{setspace}
\usepackage[ruled,vlined,linesnumbered]{algorithm2e}
\usepackage{pdfcomment}
\usepackage[table]{xcolor}
\usepackage{tabularx}
\usepackage{booktabs}
\usepackage{pdfcomment}
\usepackage{color}
\usepackage{multirow}
\definecolor{lightbluishgrey}{rgb}{0.9,0.91,0.95}

\newcommand{\TogRevision} [1] {{\textcolor[rgb]{0.00,0.00,0.00}{#1}}}

\usepackage{soul}

\usepackage{wrapfig}

\newcommand{\refequ}[1] {Eq.~\ref{equ:#1}}
\newcommand{\reffig}[1] {Figure~\ref{fig:#1}}

\newcommand{\reftab}[1] {Table~\ref{tab:#1}}

\newcommand{\refsec}[1] {Section~\ref{sec:#1}}

\newcommand{\refalg}[1] {Alg.~\ref{alg:#1}}

\let\mat = \mathbf
\usepackage{amsmath}
\usepackage{amssymb}    
\usepackage{cancel}
\usepackage[T1]{fontenc}

\newcommand{\R}{\mathbb{R}}
\newcommand{\vc}[1]{\mathbf{#1}}

\newcommand{\tr}{{\mathsf T}}

\renewcommand{\b}{\vc{b}}

\renewcommand{\d}{\vc{d}}

\newcommand{\n}{\vc{n}}

\newcommand{\p}{\vc{p}}
\newcommand{\q}{\vc{q}}
\renewcommand{\r}{\vc{r}}
\newcommand{\s}{\vc{s}}
\renewcommand{\t}{\vc{t}}

\newcommand{\x}{\vc{x}}
\newcommand{\y}{\vc{y}}
\newcommand{\z}{\vc{z}}
\newcommand{\A}{\mat{A}}
\newcommand{\B}{\mat{B}}

\newcommand{\F}{\mat{F}}
\newcommand{\G}{\mat{G}}

\newcommand{\I}{\mat{I}}
\newcommand{\J}{\mat{J}}

\renewcommand{\L}{\mat{L}}
\newcommand{\M}{\mat{M}}

\renewcommand{\S}{\mat{S}}

\newcommand{\argmin}{\mathop{\text{argmin }}}
\newcommand{\trace}{\mathop{\text{tr}}}

\newcommand{\cM}{{\cal{M}}}

\usepackage{siunitx}
\newcommand*\patchAmsMathEnvironmentForLineno[1]{%
\expandafter\let\csname old#1\expandafter\endcsname\csname #1\endcsname
\expandafter\let\csname oldend#1\expandafter\endcsname\csname end#1\endcsname
\renewenvironment{#1}%
{\linenomath\csname old#1\endcsname}%
{\csname oldend#1\endcsname\endlinenomath}}%
\newcommand*\patchBothAmsMathEnvironmentsForLineno[1]{%
\patchAmsMathEnvironmentForLineno{#1}%
\patchAmsMathEnvironmentForLineno{#1*}}%
\AtBeginDocument{%
\patchBothAmsMathEnvironmentsForLineno{equation}%
\patchBothAmsMathEnvironmentsForLineno{align}%
\patchBothAmsMathEnvironmentsForLineno{flalign}%
\patchBothAmsMathEnvironmentsForLineno{alignat}%
\patchBothAmsMathEnvironmentsForLineno{gather}%
\patchBothAmsMathEnvironmentsForLineno{multline}%
\patchBothAmsMathEnvironmentsForLineno{eqnarray}%
}

\usepackage{graphicx}

\usepackage{wrapfig}

\makeatletter
\let\save@mathaccent\mathaccent
\newcommand*\if@single[3]{%
  \setbox0\hbox{${\mathaccent"0362{#1}}^H$}%
  \setbox2\hbox{${\mathaccent"0362{\kern0pt#1}}^H$}%
  \ifdim\ht0=\ht2 #3\else #2\fi
  }
\newcommand*\rel@kern[1]{\kern#1\dimexpr\macc@kerna}
\newcommand*\widebar[1]{\@ifnextchar^{{\wide@bar{#1}{0}}}{\wide@bar{#1}{1}}}
\newcommand*\wide@bar[2]{\if@single{#1}{\wide@bar@{#1}{#2}{1}}{\wide@bar@{#1}{#2}{2}}}
\newcommand*\wide@bar@[3]{%
  \begingroup
  \def\mathaccent##1##2{%
    \let\mathaccent\save@mathaccent
    \if#32 \let\macc@nucleus\first@char \fi
    \setbox\z@\hbox{$\macc@style{\macc@nucleus}_{}$}%
    \setbox\tw@\hbox{$\macc@style{\macc@nucleus}{}_{}$}%
    \dimen@\wd\tw@
    \advance\dimen@-\wd\z@
    \divide\dimen@ 3
    \@tempdima\wd\tw@
    \advance\@tempdima-\scriptspace
    \divide\@tempdima 10
    \advance\dimen@-\@tempdima
    \ifdim\dimen@>\z@ \dimen@0pt\fi
    \rel@kern{0.6}\kern-\dimen@
    \if#31
      \overline{\rel@kern{-0.6}\kern\dimen@\macc@nucleus\rel@kern{0.4}\kern\dimen@}%
      \advance\dimen@0.4\dimexpr\macc@kerna
      \let\final@kern#2%
      \ifdim\dimen@<\z@ \let\final@kern1\fi
      \if\final@kern1 \kern-\dimen@\fi
    \else
      \overline{\rel@kern{-0.6}\kern\dimen@#1}%
    \fi
  }%
  \macc@depth\@ne
  \let\math@bgroup\@empty \let\math@egroup\macc@set@skewchar
  \mathsurround\z@ \frozen@everymath{\mathgroup\macc@group\relax}%
  \macc@set@skewchar\relax
  \let\mathaccentV\macc@nested@a
  \if#31
    \macc@nested@a\relax111{#1}%
  \else
    \def\gobble@till@marker##1\endmarker{}%
    \futurelet\first@char\gobble@till@marker#1\endmarker
    \ifcat\noexpand\first@char A\else
      \def\first@char{}%
    \fi
    \macc@nested@a\relax111{\first@char}%
  \fi
  \endgroup
}
\makeatother

\def\BibTeX{{\rm B\kern-.05em{\sc i\kern-.025em b}\kern-.08emT\kern-.1667em\lower.7ex\hbox{E}\kern-.125emX}}

\title{Towards Real-time Simulation of Hyperelastic Materials}

\author{Tiantian Liu \\
        University of Pennsylvania  \and
        Sofien Bouaziz \\
        EPFL \and
        Ladislav Kavan \\
        University of Utah
       }

\pdfauthor{Tiantian Liu, Sofien Bouaziz, Ladislav Kavan}


\begin{document}


 \teaser{
   \vspace{-1mm}
   \includegraphics[width=7.0in]{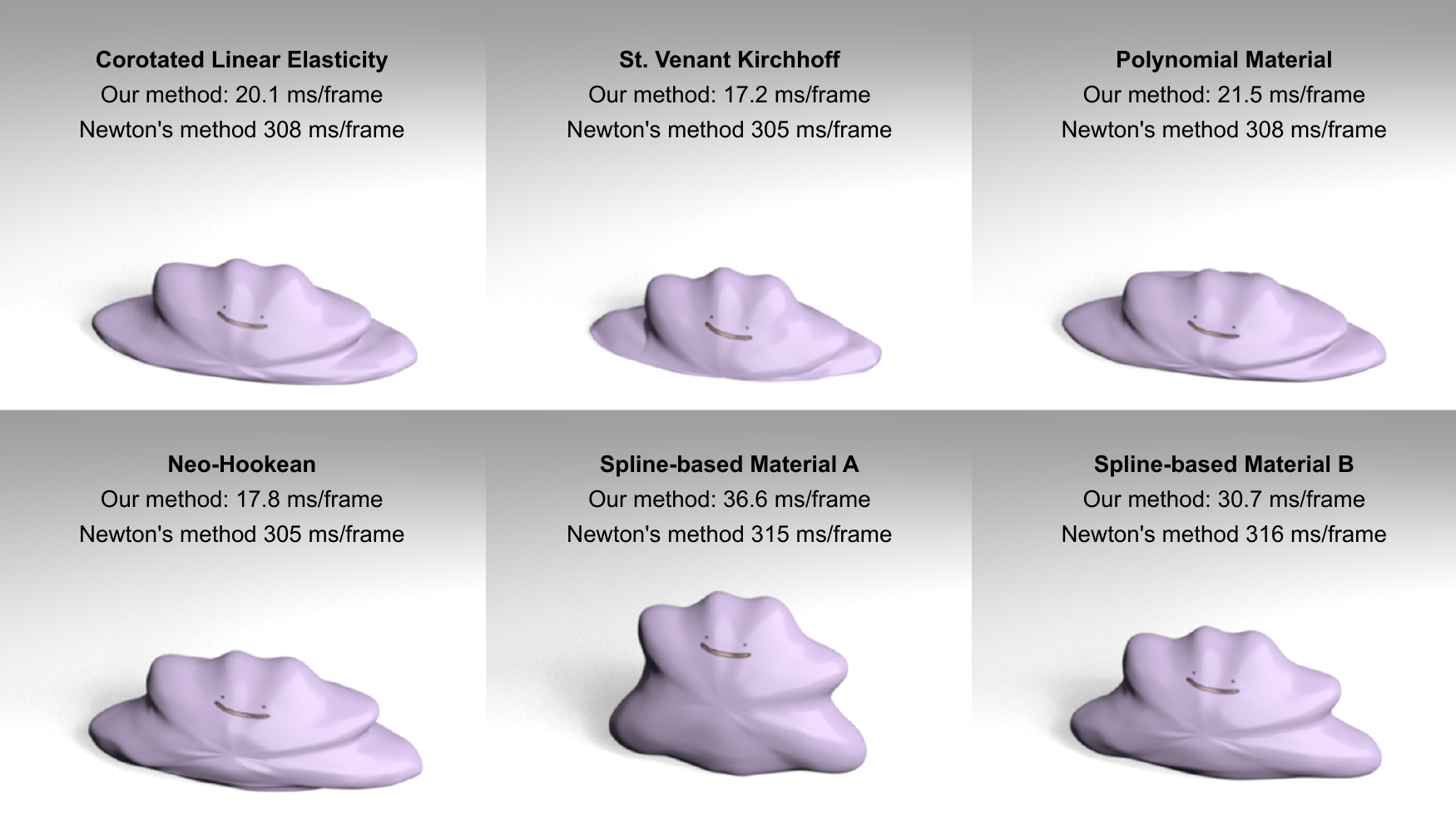}
   \vspace{-5mm}
   \caption{\label{fig:teaser}Our method enables fast simulation of many different types of hyperelastic
   materials. Compared to the commonly-applied Newton's method, our method is about 10 times
   faster, while achieving even higher accuracy and being simpler to implement.
   The Polynomial and Spline-based materials are models recently introduced by Xu et al.~\protect\shortcite{xu2015nonlinear}.
   Spline-based material A is a modified Neo-Hookean material with stronger resistance to compression;
   spline-based material B is a modified Neo-Hookean material with stronger resistance to tension.
   }
   \vspace{-2mm}
 }

\maketitle

\begin{abstract}
\vspace{-2mm}
\TogRevision{We present a new method for real-time physics-based simulation supporting many different types
of hyperelastic materials.} Previous methods such as Position Based or Projective Dynamics are
fast, but support only limited selection of materials; even classical materials such as the
Neo-Hookean elasticity are not supported. Recently, Xu et al. [2015] introduced new ``spline-based
materials'' which can be easily controlled by artists to achieve desired animation effects.
Simulation of these types of materials currently relies on Newton's method, which is slow, even
with only one iteration per timestep. In this paper, we show that Projective Dynamics can be
interpreted as a quasi-Newton method. This insight enables very efficient simulation of a large
class of hyperelastic materials, including the Neo-Hookean, spline-based materials, and others. The
quasi-Newton interpretation also allows us to leverage ideas from numerical optimization. In
particular, we show that our solver can be further accelerated using L-BFGS updates (Limited-memory
Broyden-Fletcher-Goldfarb-Shanno algorithm). Our final method is typically more than 10 times
faster than one iteration of Newton's method without compromising quality. In fact, our result is
often more accurate than the result obtained with one iteration of Newton's method. Our method is
also easier to implement, implying reduced software development costs.

\end{abstract}
\vspace{-2mm}


\keywordlist
\vspace{-0.1cm}


\section{Introduction}

\TogRevision{Physics-based animation is an important tool in computer graphics even though creating
visually compelling simulations often requires a lot of patience. Waiting for results is not an
option in real-time simulations, which are necessary in applications such as computer games and
training simulators, e.g., surgery simulators. Previous methods for real-time physics such as
Position Based Dynamics \cite{muller-position} or Projective Dynamics \cite{Bouaziz2014Projective}
have been successfully used in many applications and commercial products, despite the fact that
these methods support only a restricted set of material models.} Even classical models from
continuum mechanics, such as the Neo-Hookean, St. Venant-Kirchoff, or Mooney-Rivlin materials, are
not supported by Projective Dynamics. We tried to emulate their behavior with Projective Dynamics,
but despite our best efforts, there are still obvious visual differences when compared to
simulations with the original non-linear materials.

The advantages of more general material models were nicely demonstrated in the recent work of Xu et
al.~\shortcite{xu2015nonlinear}, who proposed a new class of spline-based materials particularly
suitable for physics-based animation. Their user-friendly spline interface enables artists to
easily modify material properties in order to achieve desired animation effects. However, their
system relies on Newton's method, which is slow, even if the number of Newton's iterations per
frame is limited to one. Our method enables fast simulation of spline-based materials, combining
the benefits of artist-friendly material interfaces with the advantages of fast simulation, such as
rapid iterations and/or higher resolutions.

Physics-based simulation can be formulated as an optimization problem where we minimize a
multi-variate function $g$. Newton's method minimizes $g$ by performing descent along direction
$-(\nabla^2 g)^{-1} \nabla g$, where $\nabla^2 g$ is the Hessian matrix, and $\nabla g$ is the
gradient. One problem of Newton's method is that the Hessian $\nabla^2 g$ can be indefinite, in
which case the Newton's direction could erroneously \textit{increase} $g$. This undesired behavior
can be prevented by so-called ``definiteness fixes''
\cite{Teran:2005:RQF:1073368.1073394,nocedal2006book}. While definiteness fixes require some
computational overheads, the slow speed of Newton's method is mainly caused by the fact that the
Hessian changes at every iteration, i.e., we need to solve a \textit{new} linear system for every
Newton step.

The point of departure for our method is the insight that Projective Dynamics can be interpreted as
a special type of quasi-Newton method. In general, quasi-Newton methods \cite{nocedal2006book} work
by replacing the Hessian $\nabla^2 g$ with a linear operator $\A$, where $\A$ is positive definite
and solving linear systems $\A \x = \b$ is fast. The descent directions are then computed as
$-\A^{-1} \nabla g$ (where the inverse is of course not explicitly evaluated, in fact, $\A$ is
often not even represented with a matrix). The trade-off is that if $\A$ is a poor approximation of
the Hessian, the quasi-Newton method may converge slowly. \TogRevision{Unfortunately, coming up with an
effective approximation of the Hessian is not easy. We tried many previous quasi-Newton methods,
but even after boosting their performance with L-BFGS \cite{nocedal2006book}, we were unable to
obtain an effective method for real-time physics. We show that Projective Dynamics can be
re-formulated as a quasi-Newton method with some remarkable properties, in particular, the
resulting $\A_\mathtt{our}$ matrix is \textit{constant} and \textit{positive definite}.} This
re-formulation enables us to generalize the method to hyperelastic materials not supported by
Projective Dynamics, such as the Neo-Hookean or spline-based materials. Even though the resulting
solver is slightly more complicated than Projective Dynamics (in particular, we must employ a line
search to ensure stability), the computational overhead required to support more general materials
is rather small.

The quasi-Newton formulation also allows us to further improve convergence of our solver. We
propose using L-BFGS, which uses curvature information estimated from a certain number of previous
iterates to improve the accuracy of our Hessian approximation $\A_\mathtt{our}$. Adding the L-BFGS
Hessian updates introduces only a small computational overhead while accelerating the convergence
of our method. \TogRevision{However, this is not a silver bullet, because the performance of L-BFGS highly
depends on the quality of the initial Hessian approximation. With previous quasi-Newton methods, we
observed rather disappointing convergence properties (see \reffig{convergence_different_methods}).
However, the combination of our Hessian approximation $\A_\mathtt{our}$ with L-BFGS is quite
effective and can be interpreted as a generalization of the recently proposed Chebyshev
Semi-Iterative method for accelerating Projective Dynamics \cite{wang2015chebyshev}.}

The L-BFGS convergence boosting is compatible with our first contribution, i.e., fast simulation of
complex non-linear materials. Specifically, we can simulate any materials satisfying the
Valanis-Landel assumption \cite{valanis1967strain} which includes many classical materials, such as
St. Venant-Kirchhoff, Neo-Hookean, Mooney-Rivlin, and also the recently proposed spline-based
materials \cite{xu2015nonlinear} (none of which is supported by Projective Dynamics). In summary,
our final method achieves faster convergence than Projective Dynamics while being able to simulate
a large variety of hyperelastic materials.

\section{Related Work}
\vspace{-2mm}

The work of Terzopoulos et al.~\shortcite{terzopoulos1987elastically} pioneered physics-based
animation, nowadays an indispensable tool in feature animation and visual effects. Real-time
physics became widespread only more recently, with first success stories represented by real-time
rigid body simulators, commercially offered by companies such as Havok since early 2000s. Fast
simulation of deformable objects is more challenging because they feature many more degrees of
freedom than rigid bodies. Fast simulations of deformable objects using shape matching
\cite{muller2005meshless,rivers-fastLSM} paved the way towards more general Position Based Dynamics
methods \cite{muller-position,stam-nucleus}. The past decade witnessed rapid development of
Position Based methods, including improvements of the convergence
\cite{muller2008hierarchical,kim2012long}, robust simulation of elastic models
\cite{muller2011solid}, generalization to fluids \cite{macklin2013position} and continuum-based
materials \cite{muller2014strain,bender2014position}, unified solvers including multiple phases of
matter \cite{macklin2014unified}, and most recently, methods to avoid element inversion
\cite{muller2015air}. We refer to a recent survey \cite{bender2014survey} for a more detailed
summary of Position Based methods.

\vspace{-1mm}

A new interpretation of Position Based methods was offered by Liu et al.~\shortcite{liu2013fast},
observing that Position Based Dynamics can be interpreted as an approximate solver for Implicit Euler
time-stepping. The same paper introduces a fast local/global solver for mass-spring systems
integrated using Implicit Euler. This method was later generalized to Projective Dynamics
\cite{Bouaziz2014Projective} by combining the ideas of \cite{liu2013fast} with a shape editing
system ``Shape-Up'' \cite{bouaziz2012shape}. Recently, a Chebyshev Semi-Iterative method
\cite{wang2015chebyshev} has been proposed to accelerate convergence of Projective Dynamics, while
exploring also highly parallel GPU implementations of real-time physics.

\vspace{-1mm}

Multi-grid methods represent another approach to accelerate physics-based simulations
\cite{georgii2006multigrid,muller2008hierarchical,wang2010multi,mcadams2011efficient,Tamstorf:2015:SAM:2816795.2818081}.
Multi-grid methods are attractive especially for highly detailed meshes where sparse direct solvers
become hindered by high memory requirements. However, constructing multi-resolution data structures
and picking suitable parameters is not a trivial task. Another way to speed up FEM is by using
subspace simulation where the nodal degrees of freedom are replaced with a low-dimensional linear
subspace~\cite{Barbic2005,An2008,Li2014}. These methods can be very efficient; however,
deformations that were not accounted for during the subspace construction may not be well
represented. A variety of approaches have been designed to address this limitation while trying to
preserve efficiency~\cite{Harmon2013,Teng2014,Teng2015}. Simulating at coarser resolutions is also
possible, while crafting special data-driven materials which avoid the loss of accuracy typically
associated with lower resolutions \cite{chen15ddfem}.

\vspace{-1mm}

The concept of constraint projection, which appears in both Position Based and Projective Dynamics,
is also central to the Fast Projection method \cite{Goldenthal2007} and strain-limiting techniques
\cite{thomaszewski2009continuum,narain2012adaptive}. The Fast Projection method and Position Based
Dynamics formulate physics simulation as a constrained optimization problem that is solved by
linearizing the constraints in the spirit of sequential quadratic programming~\cite{macklin2014unified}.
The resulting Karush-Kuhn-Tucker (KKT) equation system is then solved using a direct solver~\cite{Goldenthal2007}
or an iterative method such as Gauss-Seidel~\cite{muller-position,stam-nucleus,Fratarcangeli2015}, Jacobi~\cite{macklin2013position}, or their under/over-relaxation counterparts~\cite{macklin2014unified}.
By using a constrained optimization formulation the Fast Projection method and Position Based
Dynamics are designed for solving infinitely stiff systems but are not appropriate to handle soft materials.
This problem can be overcome by regularizing the KKT system~\cite{Servin2006,tournier2015stable},
leading to approaches that can accurately handle extremely high tensile forces (e.g.,
string of a bow) but also support soft (compliant) constraints. However, these methods are slower
than Projective Dynamics because a new linear system has to be solved at each iteration.

\vspace{-1mm}

\TogRevision{The idea of quasi-Newton methods in elasticity is not new and has been studied long time before
real-time simulations were feasible.
Several research papers have been done to accelerate Newton's method in FEM simulations by updating the Hessian approximation only once every frame~\cite{bathe1980some,fish1995efficient}. However, even one Hessian update is usually so computationally expensive that can not fit into the computing time limit of real-time applications.
Deuflhard~\shortcite{deuflhard2011newton} minimizes the number of Hessian factorizations by
carefully scheduled Hessian updates. But the update rate will heavily depend on the deformation. A
good Hessian approximation suitable for realtime applications should be easy to refactorize or
capable of prefactorization. One straightforward constant approximation which is good for
prefactorization is the Hessian evaluated at the rest-pose (undeformed configuration). The
rest-pose is positive semi-definite and its use at any configuration enables pre-factorization.
Unfortunately, the actual Hessian of deformed configurations is often very different from the
rest-pose Hessian and this approximation is therefore not satisfactory for larger deformations
\cite{Muller2002vertexwarp}.}

\TogRevision{To improve upon this, M\"{u}ller et al.~\shortcite{Muller2002vertexwarp} introduced
per-vertex ``stiffness warping'' of the rest-pose Hessian, which is more accurate and can still
leverage pre-factorized rest-pose Hessian. Unfortunately, the per-vertex stiffness warping approach
can introduce non-physical ghost forces which violate momentum conservation and can lead to
instabilities \cite{Muller2004elementwarp}. This problem was addressed by {\em per-element}
stiffness warping \cite{Muller2004elementwarp} which avoids the ghost forces but, unfortunately,
the per-element-warped stiffness matrices need to be re-factorized, introducing computational
overheads which are prohibitive in real-time simulation.
%
%
For corotated elasticity, Hecht et al.~\shortcite{Hecht:2012:USC} proposed an improved method which
can re-use previously computed Hessian factorization. Specifically, the sparse Cholesky factors are
updated only {\em when} necessary and also only {\em where} necessary. This spatio-temporal staging
of Cholesky updates improves run-time performance, however, the Cholesky updates are still costly
and their scheduling can be problematic especially in real-time applications, which require
approximately constant per-frame computing costs. Also, the frequency of Cholesky updates depends
on the simulation: fast motion with large deformations will require more frequent updates and thus
more computation, or risking ghost forces and potential instabilities. Neither is an option in
real-time simulators.}

Our re-formulation of Projective Dynamics as a quasi-Newton method reveals relationships to so
called ``Sobolev gradient methods'', which have been studied since the 1980s in the continuous
setting \cite{neuberger1983steepest}; see also the more recent monograph
\cite{neuberger2009sobolev}. The idea of quasi-Newton methods appears already in
\cite{desbrun1999interactive,hauth2001high} in the context of mass-spring systems and, more recently, in
\cite{martin2013efficient} in the context of geometry processing. Martin et
al.~\shortcite{martin2013efficient} also propose multi-scale extensions and discuss an application
in physics-based simulation, but consider only the case of thin shells and their numerical method
alters the physics of the simulated system.
Quasi-Newton methods are also useful in situations
where computation of the Hessian would be expensive or impractical \cite{nocedal2006book}. In
character animation, Hahn et al.~\shortcite{Hahn2012} used BFGS to simulate physics in ``Rig
Space'', which is challenging because the rig is a black box function and its derivatives are
approximated using finite differences.

\section{Background}\label{sec:background}
\vspace{-0.15cm}

\textbf{Projective Dynamics.} We start by introducing our notation and recapitulating the key
concepts of Projective Dynamics. Let $\x \in \R^{n \times 3}$ be the current (deformed) state of
our system containing $n$ nodes, each with three spatial dimensions. Projective Dynamics requires a
special form of elastic potential energies, based on the concept of constraint projection.
Specifically, Projective Dynamics energy for element number $i$ is defined as:
\begin{equation}\label{equ:PDenergy}
E_i(\x) = \min_{\p_i \in \cM_i} \tilde{E}_i(\x, \p_i), \ \ \tilde{E}_i(\x, \z) = \| \G_i \x - \z \|^2_F
\end{equation}
where $\|\cdot\|_F$ is the Frobenius norm, $\cM_i$ is a constraint manifold, $\p_i$ is an auxiliary
``projection variable'', and $\G_i$ is a discrete differential operator represented, e.g., by a
sparse matrix. For example, if element number $i$ is a tetrahedron, $\cM_i$ is $SO(3)$, and $\vc
G_i$ is deformation gradient operator \cite{Sifakis2012FEM}, we obtain the well-known
as-rigid-as-possible material model \cite{chao2010simple}. Another elementary example is a spring,
where the element is an edge, $\cM_i$ is a sphere, and $\vc G_i$ subtracts two endpoints of the
spring. If all elements are springs, Projective Dynamics becomes equivalent to the work of Liu et
al.~\shortcite{liu2013fast}. The key property of $\vc G_i$ is that constant vectors are in its
nullspace, which makes $E_i$ translation invariant. The total energy of the system is:
\begin{equation}\label{equ:PDtotalEnergy}
E(\x) = \sum_{i} w_i E_i(\x)
\end{equation}
where $i$ indexes elements and $w_i > 0$ is a positive weight, typically defined as the product of
undeformed volume and stiffness.

{\bf Time integration.} As discussed by Martin et al.~\shortcite{martin2011example}, Backward Euler
time integration can be expressed as a minimization of:
\begin{equation}\label{equ:PD}
g(\x) = \frac{1}{2h^2}\underbrace{\trace((\x - \y)^\tr \M (\x - \y))}_{\mathrm{inertia}} + \ \!\!\!\!\underbrace{E(\vc x)}_{\mathrm{elasticity}}
\end{equation}
where $\y$ is a constant depending only on previously computed states, $\M$ is a positive definite
mass matrix (typically diagonal -- mass lumping), and $h > 0$ is the time step (we use fixed $h$
corresponding to the frame rate of $30fps$, i.e., $h = 1/30s$). The trace ($\trace$) reflects the fact that
there are no dependencies between the $x,y,z$ coordinates, which enables us to work only with $n
\times n$ matrices (as opposed to more general $3n \times 3n$ matrices). This is somewhat moot in
the context of the mass matrix $\M$, but it will be more important in the following. The constant
$\y$ is defined as $\y := 2\q_l - \q_{l-1} + h^2 \vc M^{-1} \vc f_{\text{ext}}$, where $\q_l \in
\R^{n \times 3}$ is the current state, $\q_{l-1}$ the previous state, and $\vc f_{\text{ext}} \in
\R^{n \times 3}$ are external forces such as gravity. The minimizer of $g(\x)$ will become the next
state, $\q_{l+1}$. Intuitively, the first term in \refequ{PD} can be interpreted as ``inertial
potential,'' attracting $\x$ towards $\y$, where $\y$ corresponds to state predicted by Newton's
first law -- motion without the presence of any internal forces. The second term penalizes states
$\x$ with large elastic deformations. Minimization of $g(\x)$ corresponds to finding balance
between the two terms. Note that many other implicit integration schemes can also be expressed as
minimization problems similar to \refequ{PD}. In particular, we have implemented Implicit Midpoint,
which has the desirable feature of being symplectic \cite{hairer2002book,Kharevych2006}.
Unfortunately, in our experiments we found Implicit Midpoint to be markedly less stable than
Backward Euler and, therefore, we continue to use Backward Euler despite its numerical damping.

{\bf Local/global solver.} The key idea of Projective Dynamics is to expose the auxiliary
projection variables $\p_i$, taking advantage of the special energy form according to
\refequ{PDenergy}. To simplify notation, we stack all projection variables into $\p \in \R^{c
\times 3}$ and define binary selector matrices $\S_i$ such that $\p_i = \S_i \p$. Projective
Dynamics uses the augmented objective:
\begin{equation}\label{equ:PDaug}
\tilde{g}(\x,\p) = \frac{1}{2h^2}{\trace((\x - \y)^\tr \vc \M (\x - \y))} + \sum_i w_i \tilde{E}(\x, \S_i \p)
\end{equation}
which is minimized over both $\x$ and $\p$, subject to the constraint $\p \in \cM$, where $\cM$ is
a cartesian product of the individual constraint manifolds. The optimization is solved using an
alternating (local/global) solver. In the local step, $\x$ is assumed to be fixed; the optimal $\p$
are given by projections on individual constraint manifolds, e.g., projecting each deformation
gradient (a $3 \times 3$ matrix) on $SO(3)$. In the global step, $\p$ is assumed to be fixed and we
rewrite the objective $\tilde{g}(\x,\p)$ in matrix form:
\begin{equation}\label{equ:PDmatrixForm}
\frac{1}{2h^2}{\trace((\x - \y)^\tr \vc \M (\x - \y))} +
\frac{1}{2} \trace(\x^\tr \L \x) - \trace(\x^\tr \J \p) + C
\end{equation}
where $\L := \sum w_i \G_i^\tr \G_i$, $\J := \sum w_i \G_i^\tr \S_i$, and the constant $C$ is
irrelevant for optimization. For a fixed $\p$, the minimization of $\tilde{g}(\x,\p)$ can be
accomplished by finding $\x$ with a vanishing gradient, i.e., $\nabla_\x \tilde{g}(\x,\p) = 0$.
Computing the gradient yields some convenient simplifications (the traces disappear):
\begin{equation}\label{equ:PDgrad}
\nabla_\x \tilde{g}(\x,\p) = \frac{1}{h^2} \M (\x - \y) + \L \x - \J \p
\end{equation}
Equating the gradient to zero leads to the solution:
\begin{equation}\label{equ:PDsolution}
\x^* = (\M / h^2 + \L)^{-1}(\J \p + \M \y /h^2)
\end{equation}
The matrix $\M / h^2 + \L$ is symmetric positive definite and therefore $\x^*$ is a global minimum
(for fixed $\p$). The key computational advantage of Projective Dynamics is that $\M / h^2 + \L$
does not depend on $\x$, which allows us to pre-compute and repeatedly reuse its sparse Cholesky
factorization to quickly solve for $\x^*$, which is the result after one local and global step. The
local and global steps are repeated for a fixed number of iterations (typically 10 or 20).

\section{Method}
\vspace{-0.15cm}

\begin{figure*}[htb]
\centering
\vspace{-3mm}
\includegraphics[width = \textwidth]{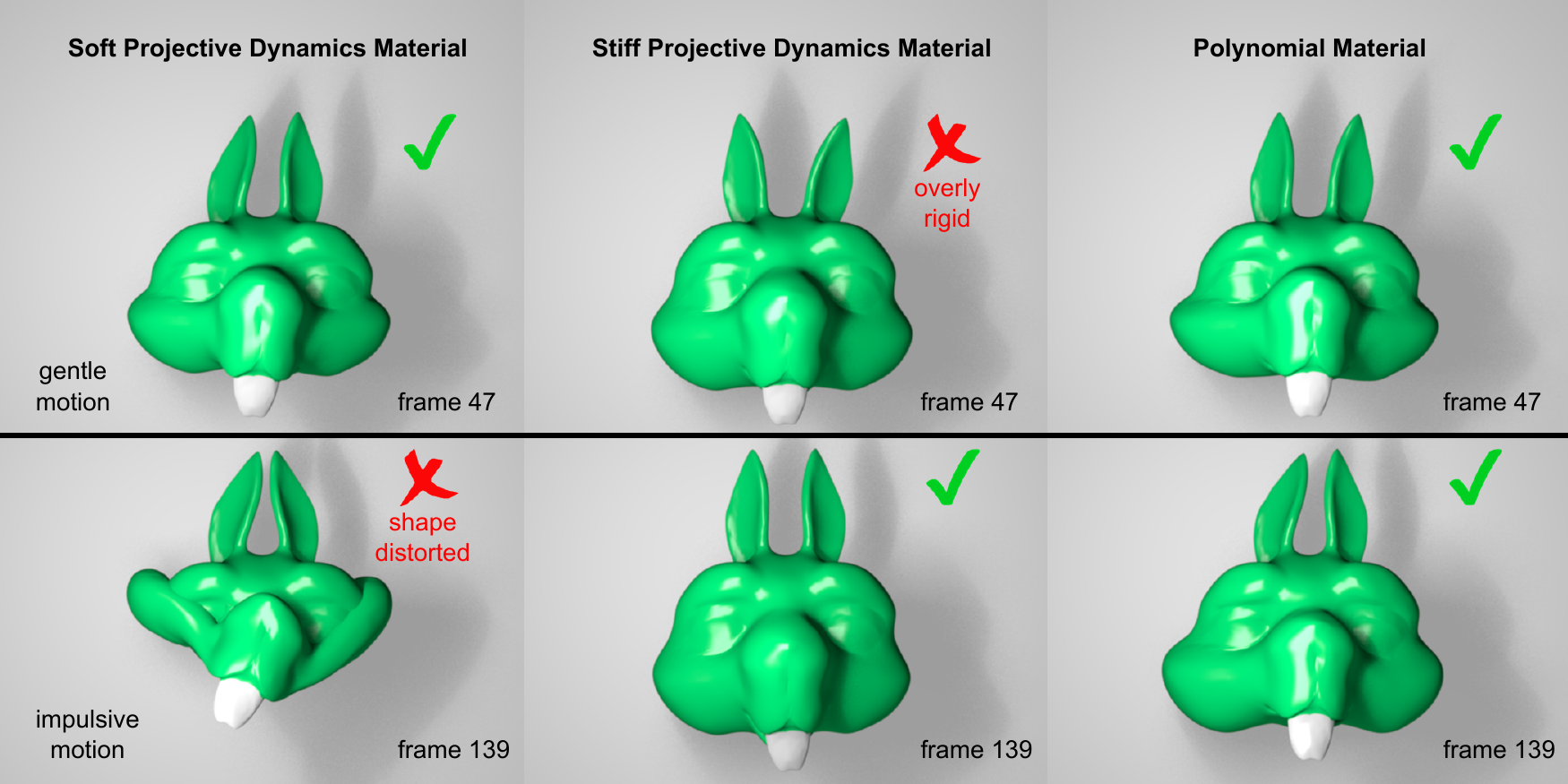}
\caption{Animating jiggly squirrel head. The squirrel head is driven by a gentle keyframed motion
in the top row, and by a faster, impulsive motion in the bottom row. Soft Projective Dynamics material
(left column) creates nice secondary motion, but does not prevent large distortions of the shape. If we stiffen
the Projective Dynamics material (middle column), we prevent the distortions,
but also kill the secondary motion. Our polynomial material $a(x) = \mu (x - 1)^4, b(x) = 0, c(x) = 0$
(right column) achieves the desired effect of jiggling without large shape distortions.}\label{fig:squirrel_head}
\vspace{-5mm}
\end{figure*}

As described in the previous section, Projective Dynamics relies on the special type of elastic
energies according to \refequ{PDenergy}. Let us now describe how Projective Dynamics can be
interpreted as a quasi-Newton method. The first step is to compute the gradient of the objective
$g(\x)$ from \refequ{PD}. The energy $E(\x)$ used in this objective contains constrained
minimization over the projection variables $\p_i \in \cM_i$ (see \refequ{PDenergy} and
\refequ{PDtotalEnergy}). Equivalently, we can interpret the $\p_i$ as functions of $\x$ realizing
the projections, according to \refequ{PDenergy}. Nevertheless, the gradient $\nabla g(\x)$ can
still be computed easily -- in fact, it is exactly equivalent to $\nabla_\x \tilde{g}(\x,\p)$ from
\refequ{PDgrad} where we assumed that $\p$ is constant. This at first surprising fact has been
observed in previous work \cite{chao2010simple,bouaziz2012shape}. Intuitively, the reason is that
if we infinitesimally perturb $\x$, its projection $\p_i(\x)$ can move only in the tangent space of
$\cM_i$ and therefore, the differential $\delta \p_i(\x)$ has no effect on $\delta \| \x - \p_i(\x)
\|^2$. As an intuitive explanation, imagine that $\x$ is a space shuttle projected to its closest
point on Earth $\p_i(\x)$; to first order, the distance of the space shuttle from Earth does not
depend on the tangent motion $\delta \p_i(\x)$. Please see the Appendix for a more formal
discussion. In summary, the gradient of \refequ{PD} is:
\begin{equation}\label{equ:PDgrad2}
\nabla {g}(\x) = \frac{1}{h^2} \M (\x - \y) + \L \x - \J \p(\x)
\end{equation}
where $\p(\x)$ is a function stacking all of the individual projections $\p_i(\x)$. Newton's method
would proceed by computing second derivatives, i.e, the Hessian matrix $\nabla^2 g(\x)$, and use it
to compute a descent direction $-(\nabla^2 g(\x))^{-1} \nabla g(\x)$. Note that definiteness fixes
may be necessary to guarantee this will really be a \textit{descent} direction
\cite{gast2015optimization}.

What happens if we modify Newton's method by using $\M / h^2 + \L$ instead of the Hessian $\nabla^2
g(\x)$? Simple algebra reveals:
\begin{equation*}\label{equ:quasiNewton}
(\M/h^2 + \L)^{-1} \nabla {g}(\x) = \x - (\M/h^2 + \L)^{-1} (\J \p(\x) + \M \y / h^2)
\end{equation*}
However, the latter term is equivalent to the result of one iteration of the local/global steps of
Projective Dynamics, see \refequ{PDsolution}. Therefore, $(\M/h^2 + \L)^{-1} \nabla {g}(\x) = \x -
\x^*$ and we can interpret $\d_{\mathrm{PD}} := -(\M/h^2 + \L)^{-1} \nabla {g}(\x)$ as a descent
direction (this time there is no need for any definiteness fixes). Projective Dynamics can be
therefore understood as a quasi-Newton method which computes the next iterate as $\x +
\d_{\mathrm{PD}}$. Typically, quasi-Newton methods use line search techniques
\cite{nocedal2006book} to find parameter $\alpha > 0$ such that $\x + \alpha \d_{\mathrm{PD}}$
reduces the objective as much as possible. However, with Projective Dynamics energies according to
\refequ{PDenergy}, the optimal value is always $\alpha = 1$.

\subsection{More general materials}\label{sec:genmat}
\vspace{-2mm}

The interpretation of Projective Dynamics as a quasi-Newton method suggests that a similar
optimization strategy might be effective for more general elastic potential energies. First, let us
focus on isotropic materials, deferring the discussion of anisotropy to \refsec{anisotropy}. The
assumption of isotropy (material-space rotation invariance) together with world-space rotation
invariance means that we can express elastic energy density function $\Psi$ as a function of
singular values of the deformation gradient \cite{irving2004invertible,Sifakis2012FEM}. In the
volumetric case, we have three singular values $\sigma_1, \sigma_2, \sigma_3 \in \R$, also known as
``principal stretches''. The function $\Psi(\sigma_1, \sigma_2, \sigma_3)$ must be invariant to any
permutation of the principal stretches, e.g., $\Psi(\sigma_1, \sigma_2, \sigma_3) = \Psi(\sigma_2,
\sigma_1, \sigma_3)$ etc. Because directly working with such functions $\Psi$ could be cumbersome,
we instead use the Valanis-Landel hypothesis \cite{valanis1967strain}, which assumes that:
\begin{equation}\label{equ:valanis-landel}
\begin{split}
\Psi(\sigma_1, \sigma_2, \sigma_3) = a(\sigma_1) + a(\sigma_2) + a(\sigma_3) + \\
b(\sigma_1 \sigma_2) + b(\sigma_2 \sigma_3) + b(\sigma_1 \sigma_3) +
c(\sigma_1 \sigma_2 \sigma_3)
\end{split}
\end{equation}
where $a, b, c: \R \rightarrow \R$. Many material models can be written in the Valanis-Landel form,
including linear corotated material \cite{Sifakis2012FEM}, St. Venant-Kirchhoff, Neo-Hookean, and
Mooney-Rivlin. The recently proposed spline-based materials \cite{xu2015nonlinear} are also based
on the Valanis-Landel assumption. How can we generalize Projective Dynamics to these types of
materials? Invoking the quasi-Newton interpretation discussed above, our method will minimize the
objective $g$ by performing descent along direction $\d(\x) := -(\M/h^2 + \L)^{-1} \nabla {g}(\x)$.
The mass matrix $\M$ and step size $h$ are defined as before, and computing $\nabla {g}(\x)$ is
straightforward. But how to define a matrix $\L$ for a given material model? This matrix can still
have the form $\L := \sum w_i \G_i^\tr \G_i$, but the question is how to choose the weights $w_i$.
In Projective Dynamics, we assumed the weights are given as $w_i = V_i k_i$, where $V_i > 0$ is
rest-pose volume of $i$-th element, and $k_i > 0$ is a stiffness parameter provided by the user. In
our case, the user instead specifies a material model according to \refequ{valanis-landel} from
which we have to infer the appropriate $k_i$ value. In the following we drop the subscript $i$ for
ease of notation.

For linear materials (Hooke's law), stiffness is given as the second derivative of elastic energy.
Therefore, it would be tempting to set $k$ equal to the second derivative of $\Psi$ at the rest
pose (corresponding to $\sigma_1 = \sigma_2 = \sigma_3 = 1$), which evaluates to $a''(1) + 2b''(1)
+ c''(1)$, regardless of whether we differentiate with respect to $\sigma_1$, $\sigma_2$, or
$\sigma_3$. Even though this method would produce suitable $k$ for some materials (such as
corotated elasticity), it does not work e.g. for a polynomial material defined as $a(x) = \mu (x -
1)^4, b(x) = 0, c(x) = 0$. Already this relatively simple material can facilitate certain animation
tasks, such as creating a cartoon squirrel head which jiggles, but does not overly distort its
shape, see \reffig{squirrel_head}. However, with this material, the second derivatives at $x = 1$
evaluates to zero regardless of the value of $\mu$, which would lead to zero stiffness which is
obviously not a good approximation. The problem is the second derivative takes into account only
infinitesimally small neighborhood of $x = 1$, i.e., the rest pose. However, we need a
\textit{single} value of $k$ which will work well in the entire range of deformations expected in
our simulations. To capture this requirement, we define an interval $[x_\mathrm{start},
x_\mathrm{end}]$ where we expect our principal stretches to be.
We consider the following stress function:
\begin{equation}\label{equ:stiffness}
\begin{split}
\left. \frac{\partial \Psi}{\partial \sigma_1} \right|_{\sigma_2 = 1, \sigma_3 = 1} = a'(\sigma_1) + 2b'(\sigma_1) + c'(\sigma_1)
\end{split}
\end{equation}
and define our $k$ as the slope of the best linear approximation of \refequ{stiffness} for
$\sigma_1 \in [x_\mathrm{start}, x_\mathrm{end}]$. Note that due to the symmetry of the
Valanis-Landel assumption, we would obtain exactly the same result if we differentiated with
respect to $\sigma_2$ or $\sigma_3$ (instead of $\sigma_1$ as above). \TogRevision{We study
different choices of $[x_\mathrm{start}, x_\mathrm{end}]$ intervals in \refsec{result}. In summary,
the results are not very sensitive on the particular choice of $x_\mathrm{start}$ and
$x_\mathrm{end}$. The key fact is that regardless of the specific setting of $x_\mathrm{start}$ and
$x_\mathrm{end}$, spatial variations of $\mu$ are correctly taken into account, i.e., softer and
stiffer parts of the simulated object will have different $\mu$ coefficients (e.g., in our squirrel
head we made the teeth more stiff). Even though all elements have the same $[x_\mathrm{start},
x_\mathrm{end}]$ interval, the resulting matrices $\L$ and $\J$ properly reflect the spatially
varying stiffness.}

\begin{algorithm}[bht]
\SetAlgoLined \SetKwFor{Loop}{loop}{times}{end}
$\x_1 := \y$; $g(\x_{1}) := \mathtt{evalObjective}(\x_{1})$\\
\For{$k = 1, \dots, \mathrm{numIterations}$}{
    $\nabla g(\x_k) := \mathtt{evalGradient}(\x_k)$ \\
    $\d(\x_{k}) := -(\M/h^2 + \L)^{-1} \nabla {g}(\x_{k})$ \\
    $\alpha := 2$ \\
    \Repeat{{$g(\x_{k+1}) \leq g(\x_{k}) + \gamma \alpha \ \mathrm{tr}((\nabla g(\x_{k}))^\tr \d(\x_{k}))$}}
    {
        $\alpha := \alpha / 2$ \\
        $\x_{k+1} := \x_{k} + \alpha \d(\x_{k})$ \\
        $g(\x_{k+1}) := \mathtt{evalObjective}(\x_{k+1})$ \\
    }
 }
\vspace{-5mm}
\caption{Quasi-Newton Solver} \label{alg:quasiNewton}
\end{algorithm}

\begin{figure}[t]
\centering
\includegraphics[width = \columnwidth]{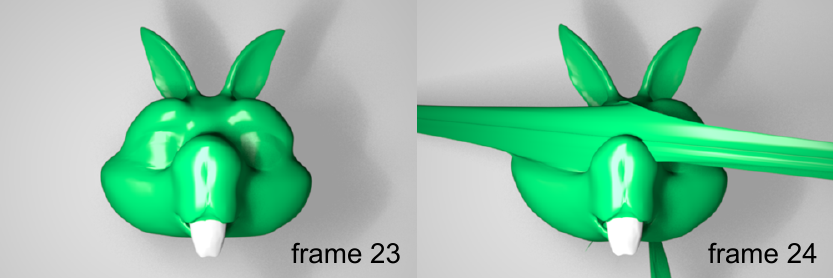}
\caption{Without line search, the squirrel head animation using our polynomial material (as
in \reffig{squirrel_head}) quickly becomes unstable.}
\label{fig:explosion}
\end{figure}

\textbf{Line search.} With Projective Dynamics materials (\refequ{PDenergy}), the line search
parameter $\alpha = 1$ is always guaranteed to decrease the objective $g$ (\refequ{PD}).
Unfortunately, this is no longer true in our generalized quasi-Newton setting, where it is easy to
find examples where $g(\x + \d(\x)) > g(\x)$, i.e., taking a step of size one actually
\textit{increases} the objective. This can lead to erroneous energy accumulation, potentially
resulting in catastrophic failure of the simulation (``explosions''), as shown in
\reffig{explosion}. Fortunately, thanks to the fact that $\M/h^2 + \L$ is positive definite,
$\d(\x)$ is guaranteed to be a descent direction. Therefore, there exists $\alpha > 0$ such that
$g(\x + \alpha \d(\x)) \leq g(\x)$ (unless we are already at a critical point $\nabla g(\x) =
\vc{0}$, at which point the optimization is finished). In fact, we can ask for even more, i.e., we
can always find $\alpha
> 0$ such that $g(\x + \alpha \d(\x)) \leq g(\x) + \gamma \alpha \trace((\nabla g(\x))^\tr \d(\x))$ for
some constant $\gamma \in (0,1)$ (we use $\gamma = 0.3$). This is known as the Armijo condition
which expresses the requirement of ``sufficient decrease'' \cite{nocedal2006book}, preventing the
line search algorithm from reducing the objective only by a negligible amount. Another requirement
for robust line search is to avoid too small steps $\alpha$ (even though they might satisfy the
Armijo condition). We tested two possibilities: Wolfe conditions, which impose an additional
``curvature condition'', and backtracking line search, which starts from large $\alpha$ and
progressively decreases it until the Armijo condition is satisfied. We found that in our setting
both approaches lead to comparable error reduction, but the backtracking line search is less
computationally expensive. Also, $\alpha = 1$ is an excellent initial guess for the backtracking
strategy. Therefore, in our final algorithm we implement the backtracking line search; after a
failed attempt, we multiply alpha by 0.5. This value worked well in our experiments, even though,
in theory, any constant $\in (0,1)$ could be used instead.

\refalg{quasiNewton} summarizes the process of computing one frame of our simulation. The outer
loop (lines 2-11) performs quasi-Newton iterations and the inner loop (lines 6-10) implements the
line search. 
%
What is the extra computational cost required to support more general materials? With Projective
Dynamics energies (\refequ{PDenergy}), we do not need the line search, because $\alpha = 1$ always
works. Indeed, if we drop the line search from \refalg{quasiNewton}, the algorithm becomes
equivalent to a generalized local/global process, as discussed in \refsec{background} (which is
unstable for non-Projective-Dynamics energies). Rejected line search attempts, i.e., additional
iterations of the line search, represent the main computational overhead of our method.
Fortunately, we found that in practical simulations the number of extra line search iterations is
relatively small. For example, in the squirrel head example in \reffig{squirrel_head} using the
polynomial material, we need only 4280 line search iterations for the entire sequence with 400
frames, 10 quasi-Newton iterations per frame, i.e., the \textit{average} number of line search iterations per quasi-Newton iteration is only $1.07$.
Even though in most cases the full step ($\alpha = 1$) succeeds, the Armijo safeguard is essential for
stability; if we drop it, the simulation can quickly explode, as shown in \reffig{explosion}.

\vspace{-1mm}
\subsection{Accelerating convergence}\label{sec:LBFGS}
\vspace{-2.5mm}

The connection between Projective Dynamics and quasi-Newton methods allows us to take advantage of
further mathematical optimization techniques. In this section, we discuss how to accelerate
convergence of our method using L-BFGS (Limited-memory BFGS). The BFGS algorithm
(Broyden-Fletcher-Goldfarb-Shanno) is one of the most popular general purpose quasi-Newton methods;
its key idea is to approximate the Hessian using curvature information calculated from previous
iterates, i.e., $\x_1, \dots, \x_{k-1}$. The L-BFGS modification means that we will use only the
most recent $m$ iterates, i.e., $\x_{k - m}, \dots, \x_{k-1}$; the rationale being that too distant
iterates are less relevant in estimating the Hessian at $\x_{k}$.

In \refalg{quasiNewton}, the matrix $\M / h^2 + \L$ in line 4 can be interpreted as our initial
approximation of the Hessian. This matrix is constant which on one hand enables its
pre-factorization, but on the other hand, $\M / h^2 + \L$ may be far from the Hessian $\nabla^2
g(\x_k)$, which is the reason for slower convergence compared to Newton's method
\cite{Bouaziz2014Projective}. L-BFGS allows us to develop a more accurate, state-dependent Hessian
approximation, leading to faster convergence without too much computational overhead (in our
experiments the overhead is typically less than 1\% of the simulation time, see \reftab{StatAll}).
The key to fast iterations of L-BFGS is the fact that the progressively updated approximate Hessian
$\A_k$ is not stored explicitly, which would require us to solve a new linear system $\A_k \d(\x_k)
= - \nabla g(\x_k)$ each iteration, implying high computational costs. Instead, L-BFGS implicitly
represents the \textit{inverse} of $\A_k$, i.e., linear operator $\B_k$ such that the desired
descent direction can be computed simply as $\d(\x_k) = - \B_k \nabla g(\x_k)$. The linear operator
$\B_k$ is not represented using a matrix (which would have been dense), but instead as a sequence
of dot products, known as the L-BFGS two-loop recursion, see \refalg{LBFGS}. For a more detailed
discussion of BFGS and its variants we refer to Chapters 6 and 7 of \cite{nocedal2006book}.

\begin{algorithm}[t]
\SetAlgoLined \SetKwFor{Loop}{loop}{times}{end}
$\q := -\nabla g(\x_k)$ \\
\For{$i = k-1, \dots, k-m$}{
$\s_i := \x_{i+1} - \x_i; \t_i := \nabla g(\x_{i+1}) - \nabla g(\x_{i})$; $\rho_i := \trace(\t_i^\tr \s_i)$ \\
 $\zeta_i := \trace(\s_i^\tr \q) / \rho_i$\\
 $\q := \q - \zeta_i \t_i$ \\
 }
$\r := \A_0^{-1} \q $ // $\A_0$ is initial Hessian approximation \\
\For{$i = k-m, \dots, k-1$}{
$\eta := \trace(\t_i^\tr \r) / \rho_i$ \\
$\r := \r + \s_i(\zeta_i - \eta)$ \\
 }
$\d(\x_k) := \r$ // resulting descent direction \\
\caption{Descent direction computation with L-BFGS} \label{alg:LBFGS}
\end{algorithm}

\refalg{LBFGS} requires us to provide an initial Hessian approximation $\A_0$, ideally such that
the linear system $\A_0 \r = \q$ can be solved efficiently (line 7). In our method, we use our old
friend: $\M / h^2 + \L$. \TogRevision{At first, it may seem the initialization of the Hessian is
perhaps not too important and the L-BFGS iterations quickly approach the exact Hessian. However,
this intuition is not true. In \refsec{result} we experiment with different possible
initializations of the Hessian and show that our particular choice of $\M / h^2 + \L$ outperforms
alternatives such Hessian of the rest-pose and many others. In short, the reason is that the L-BFGS
updates use only a very few gradient samples, which provide only a limited amount of information
about the exact Hessian. The appeal of the L-BFGS strategy is that it is very fast -- the compute
cost of the two for-loops in \refalg{LBFGS} is negligible compared to the cost of solving the
linear system in line 7 with our choice of $\A_0 = \M / h^2 + \L$. This is true even for high
values of $m$. In other words, the linear solve using $\M / h^2 + \L$ (line 7) is still doing the
``heavy lifting'', while the L-BFGS updates provide additional convergence boost at the cost of
minimal computational overheads.}

Upgrading our method with L-BFGS is simple: we only need to replace line 4 in \refalg{quasiNewton}
with a call of \refalg{LBFGS}. Note that for $m = 0$, \refalg{LBFGS} returns exactly the same
descent direction as before, i.e., $\d(\x_{k}) := -(\M/h^2 + \L)^{-1} \nabla {g}(\x_{k})$.
%
What is the optimal $m$, i.e., the size of the history window? Too small $m$ will not allow us to
unlock the full potential of L-BFGS. The main problem with too high $m$ is not the higher
computational cost of the two loops in \refalg{LBFGS}, but the fact that too distant iterates (such
as $\x_{k - 100}$) may contain information irrelevant for the Hessian at $\x_k$ and the result can
be even worse than with a shorter window. We found that $m = 5$ is typically a good value in our
experiments.

The recently proposed Chebyshev Semi-Iterative methods for Projective Dynamics
\cite{wang2015chebyshev} can also be interpreted as a special type of a quasi-Newton method which
utilizes two previous iterates, i.e., corresponding to $m = 2$. Indeed, in our experiments L-BFGS
with $m = 2$ exhibits similar convergence rate as the Chebyshev method, see
\reffig{convergence_different_methods} and further discussion in \refsec{result}.
Finally, we note that even though the Wolfe conditions are the recommended line search strategy for
L-BFGS, we did not observe any significant convergence benefit compared to our backtracking
strategy. However, evaluating the Wolfe conditions increases the computational cost per iteration
and therefore, we continue to rely on the backtracking strategy as described in
\refalg{quasiNewton}.

\begin{figure}[tb]
\centering
\includegraphics[width = \columnwidth]{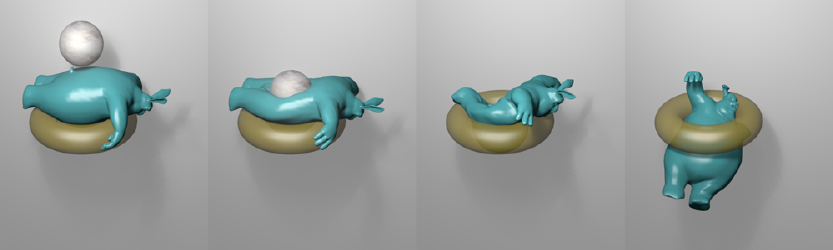}
\caption{\TogRevision{Our method is capable of simulating complex collision scenarios, such as squeezing the Big
Bunny through a torus. The Big Bunny uses corotated elasticity with $\mu=5$ and $\lambda=200$.}}
\label{fig:collision}
\vspace{-3mm}
\end{figure}

\TogRevision{
\vspace{-4mm}
\subsection{Collisions}\label{sec:collision}
\vspace{-2.5mm}
A classical approach to enforcing non-penetration constraints between deformable solids is to 1)
detect collisions and 2) resolve them using temporarily instanced repulsion springs, which bring
the volume of undesired overlaps to zero \cite{mcadams2011efficient,harmon2011interference}.
%
However, in Projective Dynamics the primary emphasis is on computational efficiency and therefore
only simplified collision resolution strategies have been proposed by Bouaziz et
al.~\shortcite{Bouaziz2014Projective}. Specifically, Projective Dynamics offers two possible
strategies. The first strategy is a two-phase method, where collisions are resolved in a separate
post-processing step using projections, similar to Position Based Dynamics. The same strategy was
employed also by Liu et al.~\shortcite{liu2013fast}. The drawback of this approach is the fact the
collision projections are oblivious to elasticity and inertia of the simulated objects. The second
approach used in Projective Dynamics is more physically realistic, but introduces additional
computational overheads. Specifically, temporarily-instanced repulsion springs are added to the
total energy. This leads to physically realistic results, but the drawback is that the matrix
$\M/h^2 + \L$ needs to be re-factorized whenever the set of repulsion springs is updated --
typically, at the beginning of each frame.}

\TogRevision{Our quasi-Newton interpretation invites a new approach to collision response which is
physically realistic, but avoids expensive re-factorizations.
%
Specifically, for each inter-penetration found by collision detection, we introduce an energy term
$E_\textrm{collision}(\x) = ( (\S \x - \t)^\tr \n )^2$, where $\S$ is a selector matrix of the
collided vertex, $\t$ is its projection on the surface and $\n$ is the surface normal.
This constraint pushes the collided vertex to the tangent plane. It is important to add this term
to our total energy $E(\x)$ only if the vertex is in collision or contact. Whenever the relative
velocity between the vertex and the collider indicates separation, the $E_\textrm{collision}(\x)$
term is discarded (otherwise it would correspond to unrealistic ``glue'' forces). This is done once
at the beginning of each iteration (just before line 3 in \refalg{quasiNewton}). The line search
(lines 6-10 of \refalg{quasiNewton}) is unaffected by these updates, i.e., the unilateral nature of
the collision constraints is handled correctly without any further processing.}

\TogRevision{ The key idea of our approach is to leverage the quasi-Newton approximation for
collision processing. In particular, we account for the $E_\textrm{collision}(\x)$ terms when
evaluating the energy and its gradients, but we ignore their contributions to the $\M/h^2 + \L$
matrix. This means that we form a somewhat more aggressive approximation of the Hessian, with the
benefit that the system matrix will never need to be re-factorized. The line search process (lines
6-10 in \refalg{quasiNewton}) guarantees that energy will decrease in spite of this more aggressive
approximation. The only trade-off we observed in our experiments is that the number of line search
iterations may increase, which is a small cost to pay for avoiding re-factorizations. We observed
that even in challenging collision scenarios, such as when squeezing a Big Bunny through a torus,
the approach behaves robustly and successfully resolves all collisions, see \reffig{collision}. }
%

\subsection{Anisotropy}\label{sec:anisotropy}
\vspace{-2mm}

\begin{figure}[htb]
\vspace{2mm}
\centering
\includegraphics[width = \columnwidth]{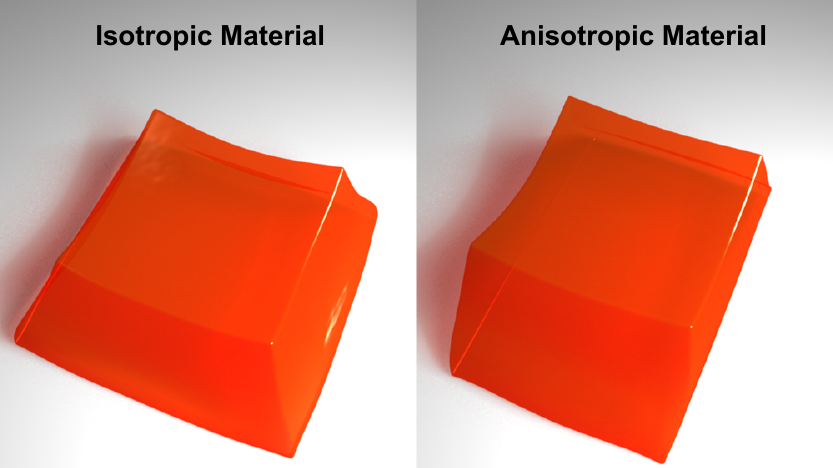}
\caption{Dropping an elastic cube on the ground. Left: deformation using isotropic elasticity (linear corotated model).
Right: the result after adding anisotropic stiffness.}\label{fig:anisotropy}
\end{figure}

Our numerical methods, including the L-BFGS acceleration, can be directly applied also to
anisotropic material models. We verified this on an elastic cube model with corotated base material
($\mu = 10$, $\lambda = 100$, referring to the notation of Sifakis and Barbi{\v
c}~\shortcite{Sifakis2012FEM}) enhanced with additional anisotropic stiffness term $
\frac{\kappa}{2}(\| \F \d \| - 1)^2$, where $\F$ is the deformation gradient and $\d$ is the
(rest-pose) direction of anisotropy. This corresponds to the directional reinforcement of the material
which is common, e.g., in biological soft tissues containing collagenous fibers. The result of our
method with $\kappa = 50$ can be seen in \reffig{anisotropy}.

\begin{figure}[bth]
\centering
\includegraphics[width = \columnwidth]{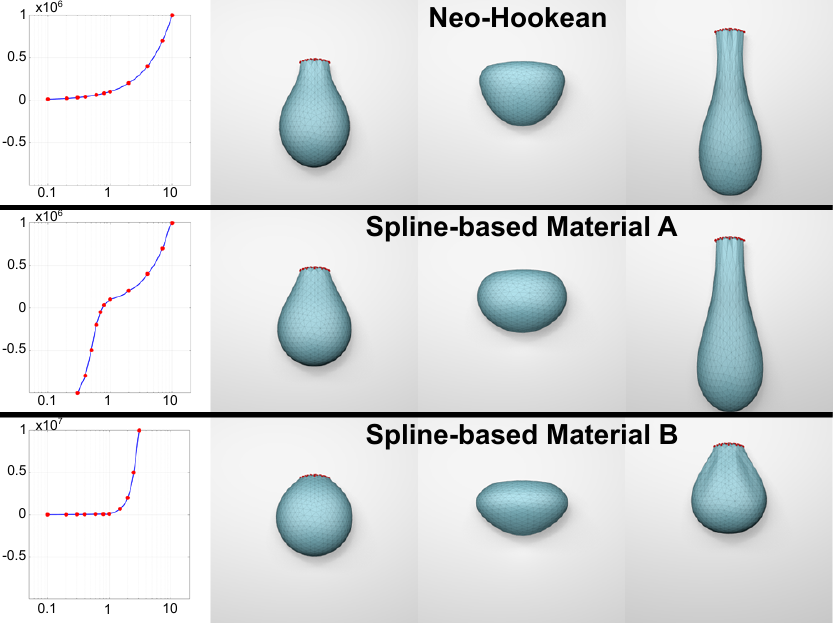}
\caption{Elastic sphere with spline-based materials \protect\cite{xu2015nonlinear}, simulated
using our method. Spline-based material A is a modified Neo-Hookean material that resists compression
more; material B is a modified Neo-Hookean material that resists tension more. The strain-stress curves
are shown on the left. }
\label{fig:spline_material}
\end{figure}

\section{Results}\label{sec:result}



\begin{table*}[t]
\centering
\begin{tabular}{|llll|rrrr|rr|}
\hline
\multirow{3}{*}{model} & \multirow{3}{*}{\#ver.} & \multirow{3}{*}{\#ele.} & \multirow{3}{*}{material model} & \multicolumn{4}{c|}{our method (10 iterations)}                                                                & \multicolumn{2}{c|}{Newton (1 iteration)}                               \\ \cline{5-10}
                         &                         &                         &                           & \multicolumn{1}{c}{linesearch}    & \multicolumn{1}{c}{L-BFGS}  & \multicolumn{1}{c}{per-frame}  & \multicolumn{1}{c|}{relative} & \multicolumn{1}{c}{per-frame}     & \multicolumn{1}{c|}{relative}     \\
                         &                         &                         &                           & \multicolumn{1}{c}{iterations} & \multicolumn{1}{c}{overhead}  & \multicolumn{1}{c}{time} & \multicolumn{1}{c|}{error}    & \multicolumn{1}{c}{time} & \multicolumn{1}{c|}{error}    \\ \hline\hline
Thin sheet               & 660                     & 1932                    & Polynomial            & 10.8                           &0.026 ms                & 4.4 ms                        & $2.7\times 10^{-8}$                     & 184 ms                            & $8.8\times 10^{-4}$                     \\
Sphere                   & 889                     & 1821                    & Spline-based A        & 24.5$^{\dagger}$\!\!\!                            &0.155 ms                 & 21.2 ms                       & $2.7\times 10^{-7}$                     & 188 ms                            & $6.9\times 10^{-4}$                     \\
Sphere                   & 889                     & 1821                    & Spline-based B        & 21.8                            &0.156 ms                 & 19.7 ms                       & $6.9\times 10^{-6}$                     & 187 ms                            & $2.5\times 10^{-4}$                     \\
\TogRevision{Shaking bar}                      & \TogRevision{574}                    & \TogRevision{1647}                   & \TogRevision{Corotated}           & \TogRevision{10.1}                            &\TogRevision{0.193 ms}                 & \TogRevision{7.2 ms}                       & \TogRevision{$1.6\times 10^{-4}$}                     & \TogRevision{171 ms}                            & \TogRevision{$4.4\times 10^{-3}$}                     \\
Ditto                    & 1454                    & 4140                    & Neo-Hookean           & 11.7                            &0.203 ms                & 17.8 ms                       & $3.0\times 10^{-5}$                     & 305 ms                            & $1.6\times 10^{-3}$                     \\
Hippo                    & 2387                    & 8406                    & Corotated             & 11.9                            &0.555 ms                      & 40.6 ms                       & $2.2\times 10^{-3}$                    & 640 ms                            & $3.7\times 10^{-2}$                     \\
Twisting bar                      & 3472                    & 10441                   & Neo-Hookean           & 10.6                            &0.945 ms                 & 45.6 ms                       & $9.4\times 10^{-5}$                     & 681 ms                            & $7.9\times 10^{-3}$                     \\
Cloth                    & 6561                    & 32160                   & Mass-Springs          & 10.0                                                            &1.20 ms                  & 42.3 ms                       & $9.3\times 10^{-4}$                     & 798 ms                            & $1.2\times 10^{-2}$                      \\
\TogRevision{Big Bunny}                 & \TogRevision{6308}                    & \TogRevision{26096}                   & \TogRevision{Corotated}            & \TogRevision{49.2$^{\ddagger}$\!\!\!}                             &\TogRevision{2.19 ms}                  & \TogRevision{623 ms}                      & \TogRevision{$9.8\times 10^{-2}$}                  & \TogRevision{2700 ms}                           & \TogRevision{$2.8\times 10^{-1}$}                    \\
Squirrel                 & 8395                    & 23782                   & Polynomial            & 10.7                             &1.41 ms                  & 153 ms                        & $8.3\times 10^{-8}$                     & 2400 ms                           & $9.1\times 10^{-6}$                     \\
Squirrel                 & 33666                   & 125677                  & Polynomial            & 10.5                             &6.38 ms                  & 706 ms                       & $1.5\times 10^{-5}$                     & 15800 ms                          & $5.4\times 10^{-5}$                     \\\hline
\end{tabular}
\caption{In all examples, we execute 10 iterations of our method per frame,
accelerated with L-BFGS with history window $m = 5$. Newton's method uses 1 iteration per frame.
The ``linesearch iterations'' reports the average number of line search iterations per frame.
The ``L-BFGS overhead'' is the runtime overhead of L-BFGS, i.e., timing of \refalg{LBFGS} without line 7
($m=5$). The reported per-frame time for
our method accounts for {\em all 10 iterations}. One iteration of our method is approximately
100 times faster than one iteration of Newton's method. We use 10 iterations of our method which reduce
the error more than one iteration of Newton's method, while being about 10 times faster.
$^{\dagger}$The higher number of line search iterations is due to the high nonlinearity of the spline-based materials and large deformations of the sphere.
$^{\ddagger}$In this case, the higher number of line search iterations is caused by nonlinearities due to
collisions (\refsec{collision}).
}
\label{tab:StatAll}
\vspace{-2mm}
\end{table*}

\begin{figure*}[htb]
\centering
\includegraphics[width = \textwidth]{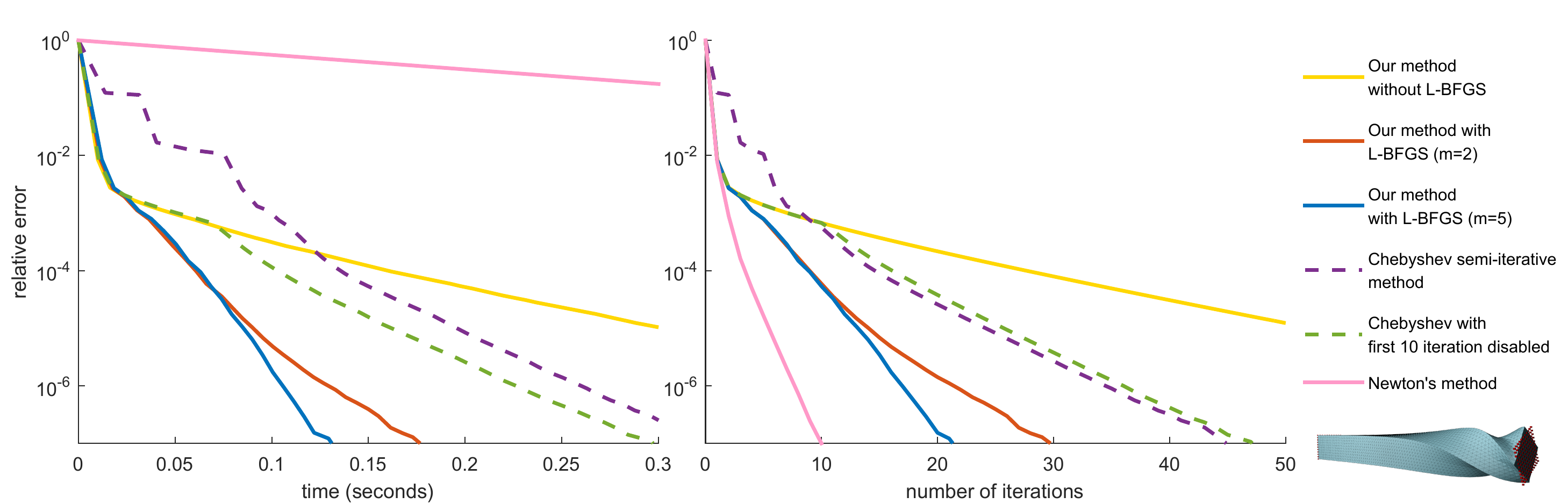}
\caption{Convergence of our method with different L-BFGS history settings, compared to Chebyshev Semi-Iterative
method and Newton's method (baseline). The model is ``Twisting bar'' with Neo-Hookean elasticity.} \label{fig:convergence_different_methods}
\vspace{-2mm}
\end{figure*}

\begin{figure*}[htb]
\centering
\includegraphics[width = \textwidth]{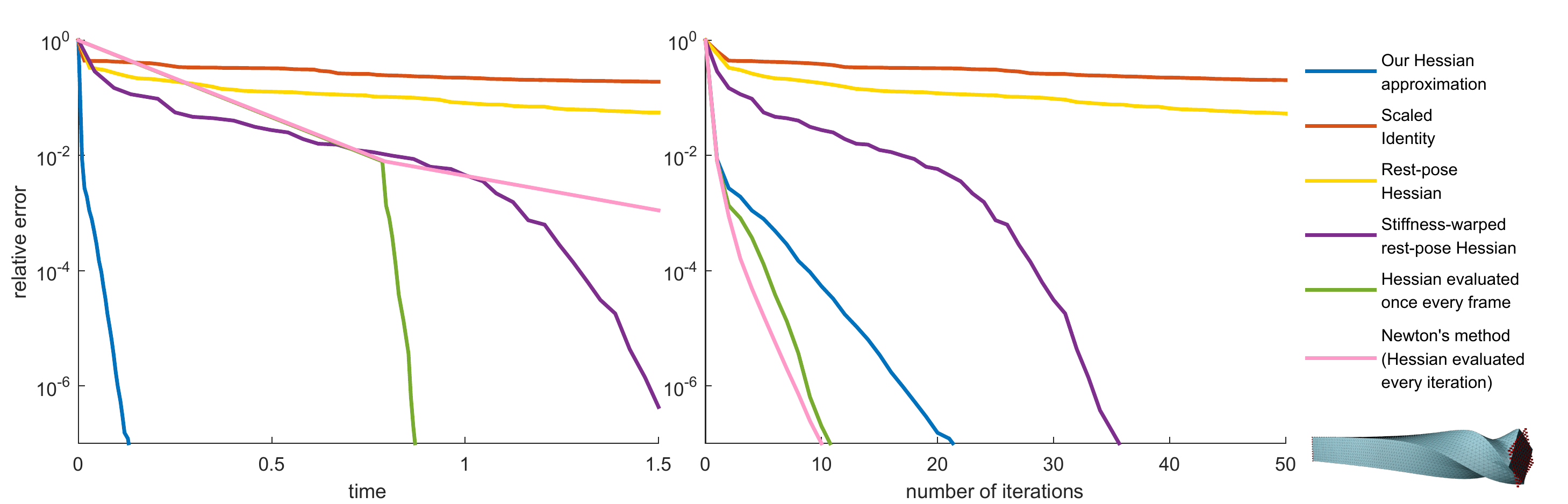}
\caption{\TogRevision{Convergence comparison of L-BFGS methods (all using $m=5$) initialized with different Hessian approximations,
along with Newton's method (baseline). The model is ``Twisting bar'' with Neo-Hookean elasticity.}} \label{fig:convergence_lbfgs}
\vspace{-2mm}
\end{figure*}


Our method supports standard elastic materials, such as corotated linear elasticity, St.
Venant-Kirchhoff and the Neo-Hookean model, see \reffig{teaser}. None of these materials is
supported by Projective Dynamics (note that Projective Dynamics supports a special sub-class of
corotated linear materials, specifically, ones with $\lambda = 0$). Our method also supports the
recently introduced spline-based materials proposed by Xu et al.~\shortcite{xu2015nonlinear}, as
shown in \reffig{teaser} and \reffig{spline_material}.

\reftab{StatAll} reports our testing scenarios and compares the run time of our method with
Newton's method, both executed on an Intel i7-4910MQ CPU at 2.90GHz. All scenarios are produced with
a fixed timestep of $1/30$ seconds. Because Newton's method is not guaranteed to work with
indefinite Hessians, we employ the standard definiteness fix \cite{Teran:2005:RQF:1073368.1073394},
i.e., we project the Hessian of each element to its closest positive definite component.
We found this method works better than other definiteness fixes, such as adding a multiple of the
identity matrix \cite{martin2011example}, which affects the \textit{entire} simulation even if
there are just a few problematic elements. The approximately 100 times faster run-time of one
iteration of our method compared to one iteration of Newton's method is due to the following facts:
1) we use pre-computed sparse Cholesky factorization, because our matrix $\M/h^2 + \L$ is constant,
2) the size of our matrix is $n \times n$, whereas the Hessian used in Newton's method is a $3n
\times 3n$ matrix, i.e., the $x,y,z$ coordinates are no longer decoupled, 3) the computation of SVD
derivatives, necessary to evaluate the Hessians of materials based on principal stretches
\cite{xu2015nonlinear}, is expensive. Note that our method is also simpler to implement, as no SVD
derivatives or definiteness fixes are necessary.

{\bf Comparison to Chebyshev Semi-Iterative method.} We compared the convergence of our method with
various lengths of the L-BFGS window to the recently introduced Chebyshev Semi-Iterative method
\cite{wang2015chebyshev}. We also plot results obtained with Newton's method as a baseline, see
\reffig{convergence_different_methods}.

Even though the Chebyshev method was originally proposed only for Projective Dynamics energies, our
generalization to arbitrary materials is compatible with the Chebyshev Semi-Iterative acceleration,
see \refalg{modifiedChebyshev}. The \refalg{modifiedChebyshev} computes a descent direction which
can be used in line 4 of \refalg{quasiNewton}. As discussed by Wang~\shortcite{wang2015chebyshev},
the Chebyshev acceleration should be disabled during the first $S$ iterations, where the
recommended value is $S = 10$. Another parameter which is essential for the Chebyshev method is an
estimate of spectral radius $\rho$, which is calculated from training simulations
\cite{wang2015chebyshev}. This parameter must be estimated carefully, because under-estimated
$\rho$ can lead to the Chebyshev method producing \textit{ascent} directions (as opposed to descent
directions). Without line search, the ascent directions manifest themselves as oscillations
\cite{wang2015chebyshev}. For the purpose of comparisons, we implemented the Chebyshev method with
direct solver which is the fastest method on the CPU \cite{wang2015chebyshev}.

\begin{algorithm}[h]
\SetAlgoLined \SetKwFor{Loop}{loop}{times}{end}
// $S$ \dots Chebyshev disabled for the first $S$ iterates, default $S$ = 10 \\
// $\rho$ \dots approximated spectral radius \\
$\q := -(\M/h^2 + \L)^{-1} \nabla {g}(\x_{k})$ \\
$\hat{\x}_{k+1} := \x_{k} + \q$\\
\lIf{$k<S$}{$\omega_{k+1}:=1$} \lIf{$k=S$}{$\omega_{k+1}:=2/(2-\rho^2)$}
\lIf{$k>S$}{$\omega_{k+1}:=4/(4-\rho^2\omega_k)$}
$\d(\x_{k}) := \omega_{k+1}(\hat{\x}_{k+1}-\x_{k-1})+\x_{k-1}-\x_{k}$ \\
\caption{Descent direction computation using Chebyshev Semi-Iterative Method
\protect\cite{wang2015chebyshev}}\label{alg:modifiedChebyshev}
\end{algorithm}

We compare the convergence of all methods using relative error, defined as:
\begin{equation}\label{equ:RelativeError}
\frac{{g}(\x_k)-{g}(\x^*)}{{g}(\x_0)-{g}(\x^*)}
\end{equation}
where $\x_0$ is the initial guess (we use $\x_0 := \y$ for all methods), $\x_k$ is the $k$-th
iterate, and $\x^*$ is the exact solution computed using Newton's method (iterated until
convergence). The decrease of relative error for one example frame is shown in
\reffig{convergence_different_methods}, where all methods are using the backtracking line search
outlined in \refalg{quasiNewton}. As expected, descent directions computed using Newton's method
are the most effective ones, as can be seen in \reffig{convergence_different_methods} (right).
However, each iteration of Newton's method is computationally expensive, and therefore other
methods can realize faster error reduction with respect to computational time, as shown in
\reffig{convergence_different_methods} (left).
All of the remaining methods are based on the constant Hessian approximation $\M/h^2 + \L$ which
leads to much faster convergence. Out of these methods, classical Projective Dynamics converges
slowest. The Chebyshev Semi-Iterative method improves the convergence; we also confirmed that
disabling the Chebyshev method during the first 10 iterations indeed helps, as recommended by Wang~
\shortcite{wang2015chebyshev}. Our method aided with L-BFGS improves convergence even further.
Already with $m=2$ (where $m$ is the size of the history window), we obtain slightly faster
convergence than with the Chebyshev method. One reason is that it is not necessary to disable
L-BFGS in the first several iterates, because L-BFGS is effective as soon as the previous iterates
become available. Also, we do not have to estimate the spectral radius which is required by the
Chebyshev method. With L-BFGS, we can also increase the history window, e.g., to $m=5$, obtaining
even more rapid convergence.

\TogRevision{ {\bf L-BFGS with different initial Hessian estimates.} Our method can be interpreted
as providing a particularly good initial estimate of the Hessian for L-BFGS. Therefore, it is
important to compare to other possible Hessian initializaitons.
In a general setting, Nocedal and Wright~\shortcite{nocedal2006book} recommend to bootstrap L-BFGS
using a scaled identity matrix:
\begin{equation}\label{equ:generalLBFGS}
\A_0 := \frac{\trace(\s_{k-1}^\tr \y_{k-1})}{\trace(\y_{k-1}^\tr \y_{k-1})} \I
\end{equation}
We experimented with this approach, but we found that our choice $\A_0 := \M/h^2 + \L$ leads to
much faster convergence, trumping the computational overhead associated with solving the
pre-factorized system $\A_0 \r = \q$ (see \reffig{convergence_lbfgs}, red graph).}

\TogRevision{Another possibility would be to set $\A_0$ equal to the rest pose Hessian, which can
of course also be pre-factorized. As shown in \reffig{convergence_lbfgs} (yellow graph), this is a
slightly better approximation than scaled identity, but still not very effective. This is because
the actual Hessian depends on world-space rotations of the model, deviating significantly from the
rest-pose Hessian. This issue was observed by M\"uller et al.~\shortcite{Muller2002vertexwarp}, who
proposed per-vertex {\em stiffness warping} as a possible remedy.
%
Per-vertex stiffness warping still allows us to leverage pre-factorization of the rest-pose Hessian
and results in better convergence than pure rest-pose Hessian, see \reffig{convergence_lbfgs}
(purple graph). However, per-vertex stiffness warping may introduce ghosts forces, because
stiffness warping uses different rotation matrices for each vertex, which means that internal
forces in one element no longer have to sum to zero. The ghost forces disappear in a fully
converged solution, however, this would require a prohibitively high number of iterations.}

\TogRevision{Yet another possibility is to completely re-evaluate the Hessian at the beginning of
each frame. This requires re-factorization, however, the remaining 10 (or so) iterations can reuse
the factorization, relying only on L-BFSG updates. When measuring convergence with respect to
number of iterations, this approach is very effective, as shown in \reffig{convergence_lbfgs}
(right, green graph). However, the cost of the initial Hessian factorization is significant, as
obvious from \reffig{convergence_lbfgs} (left, green graph). Our method uses the same Hessian
factorization for all frames, avoiding the per-frame factorization costs, while featuring excellent
convergence properties, see \reffig{convergence_lbfgs} (blue graph).}


\begin{figure}[htb]
\centering
\includegraphics[width = \columnwidth]{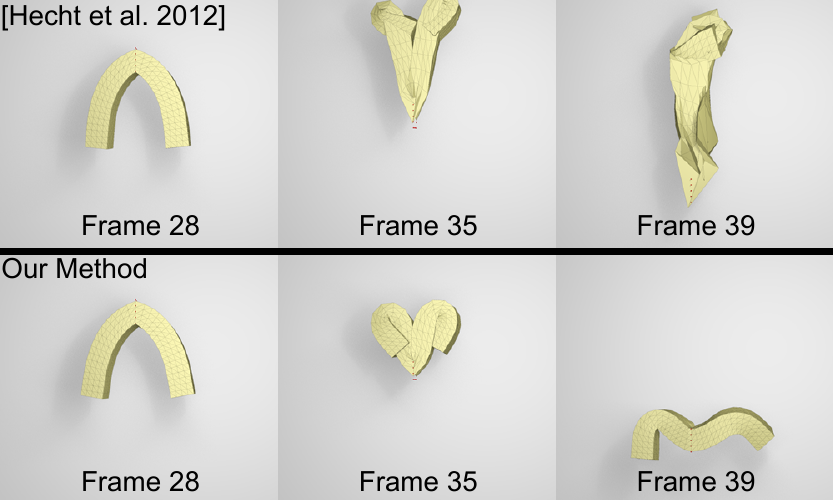}
\caption{\TogRevision{Simulation of a bar with corotated elasticity, constrained in the middle and rapidly shaken.
The method of Hecht et al.~\protect\shortcite{Hecht:2012:USC} with full Hessian updates every
{\em other} frame explodes due to large ghost forces (top).
Our method does not introduce any ghost forces and remains stable (bottom).}}
\label{fig:hecht_comparison}
\end{figure}
\TogRevision{The overheads of per-frame Hessian factorizations can be mitigated by carefully
scheduled Hessian updates. In particular, the Hessian can be reused for multiple subsequent frames
if the state is not changing too much \cite{deuflhard2011newton}. Assuming the corotated elastic
model, Hecht et al~\shortcite{Hecht:2012:USC} push this idea even further by proposing a
warp-cancelling form of the Hessian which allows not only for {\em temporal} schedule, but also for
{\em spatially} localized updates. Specifically, a nested dissection tree allows for recomputing
only parts of the mesh, which is particularly advantageous in situations where only small part of
the object is undergoing large deformations. However, the updates are still costly, and the
frequency of the updates depends on the simulation. Similarly to per-vertex stiffness warping,
insufficiently frequent update may produce ghost forces and consequent instabilities. This can be a
problem when simulating quickly moving elastic objects. To illustrate this, in
\reffig{hecht_comparison} we show a simulation of shaking an elastic bar. Even if we schedule the
Hessian updates every other frame and recompute the entire domain, this method still generates too
large ghost forces and becomes unstable. In contrast, our method remains stable and does not
require {\em any} run-time Hessian updates. }

\begin{figure}[bht]
\centering
\includegraphics[width = \columnwidth]{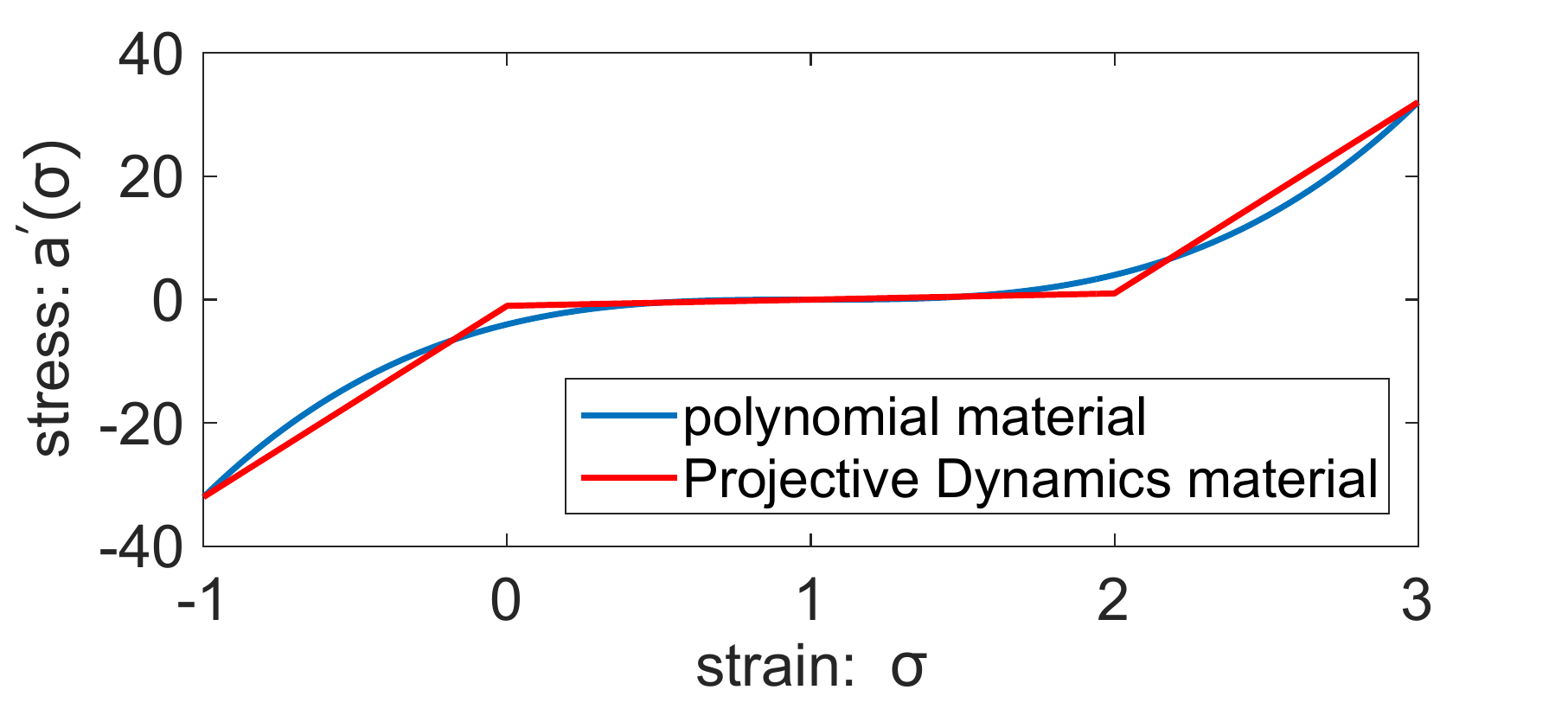}
\caption{The strain-stress curve of a polynomial material can be approximated piece-wise linearly with
two Projective Dynamics constraints.}
\label{fig:strain_stress}
\vspace{1mm}
\end{figure}

{\bf Comparison to Projective Dynamics.} One possible alternative to our method would be to apply
regular Projective Dynamics with additional strain-limiting constraints
\cite{Bouaziz2014Projective}, enabling us to construct piece-wise linear approximations of the
strain-stress curves of more general materials. We tried to use this approach to approximate the
polynomial material ($a(x) = \mu (x - 1)^4, b(x) = 0, c(x) = 0$) discussed in \refsec{genmat}, see
\reffig{strain_stress}. Even though we obtain similar overall behavior, there are two types of
artifacts associated with this approximation.
%
First, the strain-limiting constraints introduce damping when they are \textit{not} activated. This
is because the projection terms still exist in our constant matrix $\M/h^2 + \L$; if the
strain-limiting is not activated, the deformation gradients project to their current values, which
produces the undesired damping. The second problem is due to the non-smooth nature of the
piece-wise linear approximation, i.e., the stiffness of the simulated object is abruptly changed
when the strain-limiting constraints become activated. As shown in the accompanying video, our
method avoids both of these issues.

\begin{figure}[htb]
\centering
\includegraphics[width = \columnwidth]{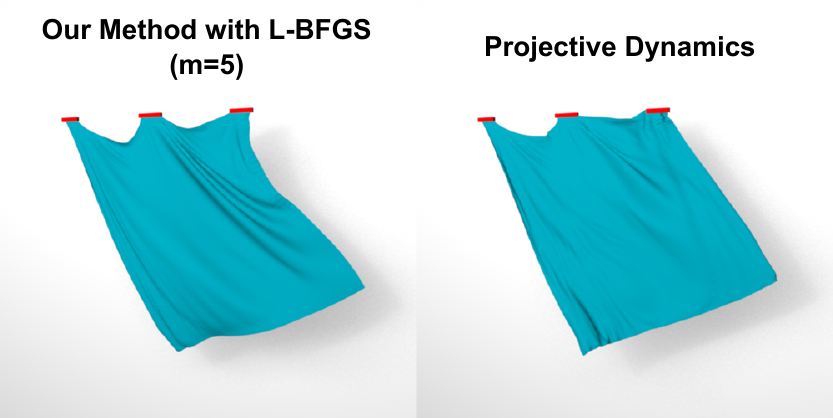}
\caption{Mass-spring system simulation using our method with L-BFGS (left) and without, i.e., using
pure Projective Dynamics (right). The L-BFGS acceleration results in more realistic wrinkles.}
\label{fig:lbfgs_visual_difference}
\end{figure}

The L-BFGS acceleration benefits also simulations which use only Projective Dynamics materials
(\refequ{PDenergy}). The most elementary example of these materials are mass-spring systems. In
\reffig{lbfgs_visual_difference}, we can see that the L-BFGS acceleration applied to a mass-spring
system simulation results in more realistic wrinkles.

\begin{figure*}[tb]
\centering
\includegraphics[width = \textwidth]{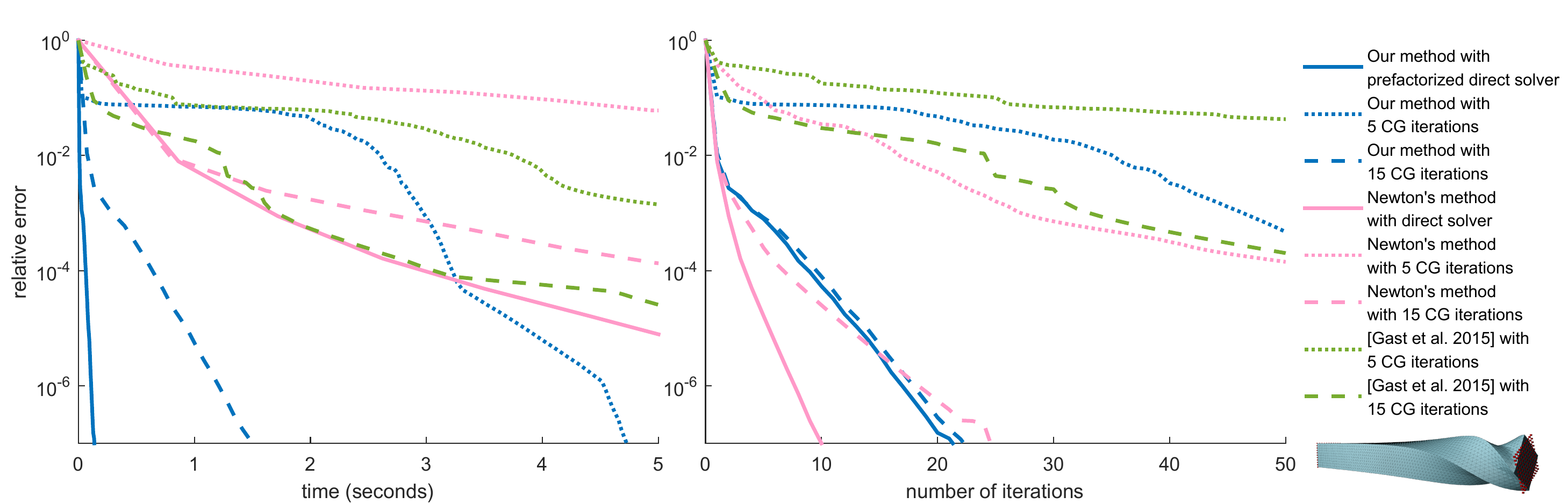}
\caption{\TogRevision{Convergence comparison of various methods using sparse direct solvers and conjugate gradients.
The model is ``Twisting bar'' with Neo-Hookean elasticity.}
} \label{fig:convergence_different_solvers}
\end{figure*}

\begin{figure}[htb]
\centering
\includegraphics[width = \columnwidth]{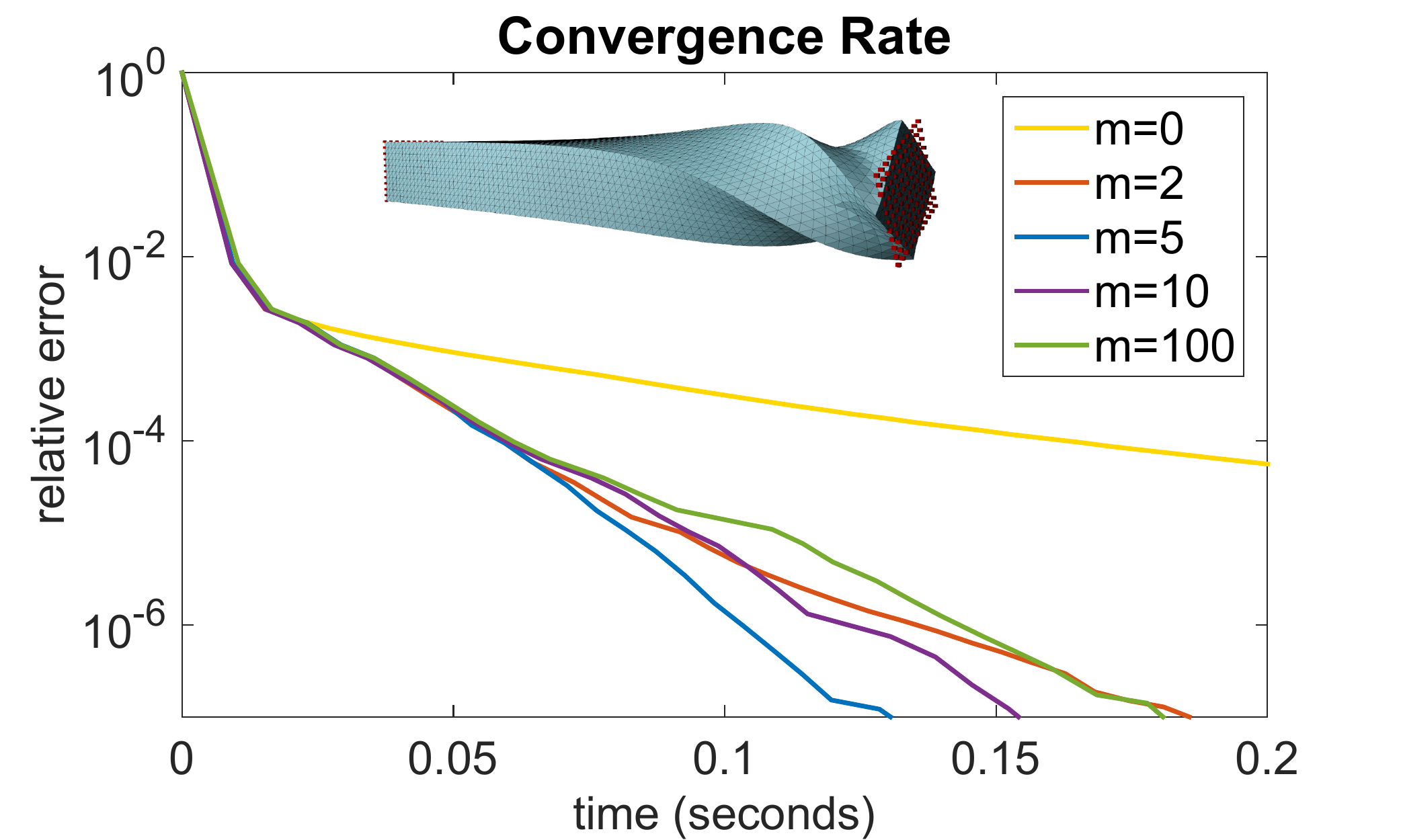}
\caption{Comparison of L-BFGS convergence rate with different history window sizes ($m$).}
\label{fig:convergence_different_history_windows}
\end{figure}

{ {\bf Choice of L-BFGS history window size.} Which history window ($m$) is the best? We
experimented with different values of $m$, see \reffig{convergence_different_history_windows}. Too
large $m$ takes into account too distant iterates which can lead to worse approximation of the
Hessian. In \reffig{convergence_different_history_windows}, we see the optimal value is $m=5$,
which is also our recommended default setting. However, it is comforting that the algorithm is not
particularly sensitive to the setting of $m$ -- even large values such as $m=100$ produce only
slightly worse convergence. In \reffig{convergence_different_methods} we can notice that the
convergence rate of the Chebyshev method is similar to our method with L-BFGS using $m=2$. We
believe this is not a coincidence, because the Chebyshev method uses \textit{two} previous
iterates, just like L-BFGS with $m=2$. }

\begin{figure}[htb]
\centering
\includegraphics[width = \columnwidth]{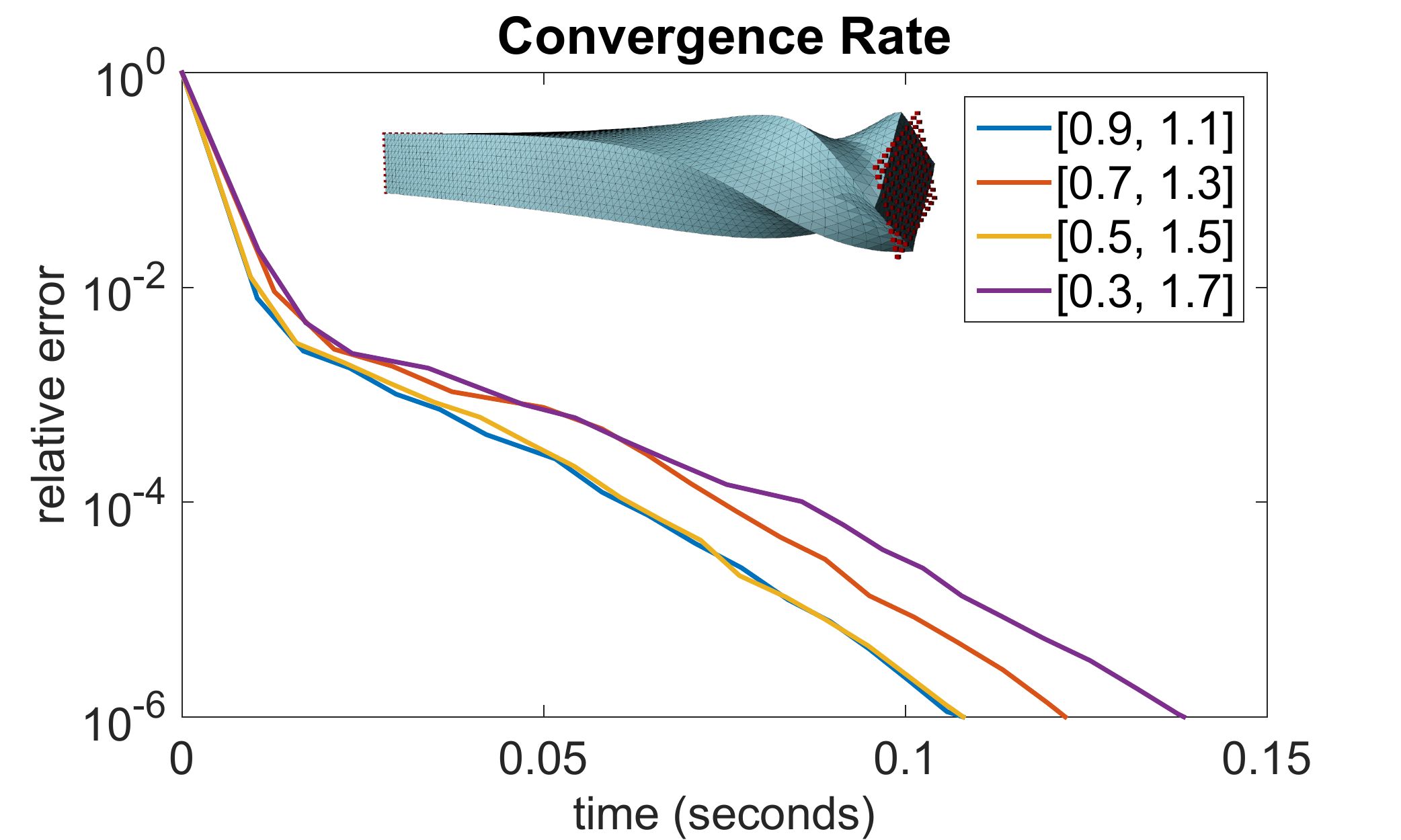}
\caption{\TogRevision{The convergence rate for different stiffness parameters chosen from different regions.}}
\label{fig:convergence_different_regions}
\end{figure}

\TogRevision{{\bf Choice of stiffness parameters.} As discussed in \refsec{genmat}, we
\refequ{stiffness} and define our stiffness parameter $k$ as the slope of the best linear
approximation of \refequ{stiffness} for $\sigma_1 \in [x_\mathrm{start}, x_\mathrm{end}]$. What is
the best $[x_\mathrm{start}, x_\mathrm{end}]$ interval to use? In the limit, with
$[x_\mathrm{start}, x_\mathrm{end}] \rightarrow [1,1]$, our $k$ would converge to the second
derivative. However, a finite interval $[x_\mathrm{start}, x_\mathrm{end}]$ guarantees that our $k$
is meaningful even for materials such as the polynomial material $a(x) = \mu (x - 1)^4, b(x) = 0,
c(x) = 0$; in this case, we obtain a $k$ which depends linearly on $\mu$.
We argue the convergence of our algorithm is not very sensitive to a particular choice of the
$[x_\mathrm{start}, x_\mathrm{end}]$ interval. In \reffig{convergence_different_regions}, we show
convergence graphs of a twisting bar with Neo-Hookean material using different intervals to compute
the stiffness parameter $k$. Although Neo-Hookean material is highly non-linear, the convergence
rates for different interval choices are quite similar. Therefore, we decided not to investigate
more sophisticated strategies and we set $x_\mathrm{start} = 0.5, x_\mathrm{end} = 1.5$ in all of
our simulations.}

\TogRevision{ {\bf Comparison with iterative solvers.}\label{sec:directVsIterative} Sparse
iterative solvers do not require expensive factorizations and are therefore attractive in
interactive applications. A particularly popular iterative method are Conjugate gradients (CG)
\cite{shewchuk1994introduction}. An additional advantage is that CG can be implemented in a
matrix-free fashion, i.e., without explicitly forming the sparse system matrix. Gast et
al.~\shortcite{gast2015optimization} further accelerate the CG solver used in Newton's method by
proposing a CG-friendly definiteness fix. Specifically, the CG iterations are terminated whenever
the maximum number of iteration is reached or indefiniteness of the Hessian matrix is detected.}

\TogRevision{
While iterative methods can be the only possible choice in high-resolution simulations (e.g., in
scientific computing), in real-time simulation scales, sparse direct solvers with pre-computed
factorization are hard to beat, as we show in \reffig{convergence_different_solvers}. Specifically,
we test Newton's method with linear systems solved using CG with 5 and 15 iterations, using Jacobi
preconditioner. Even with 15 CG iterations, the accuracy is still not the same as with the direct
solver the computational cost becomes high. If we use only five CG iterations the running time
improves, but the convergence rate suffers because the descent directions are not sufficiently
effective. The method of Gast et al.~\shortcite{gast2015optimization} initially outperforms Newton
with CG, however, the convergence slows down in subsequent iterations. We also tried to apply CG to
our method, in lieu of the direct solver. With 15 CG iterations the convergence is competitive,
however, the CG solver is slower.
}

\begin{figure}[htb]
\centering
\includegraphics[width = \columnwidth]{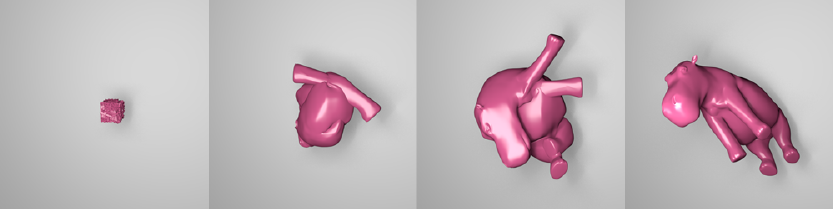}
\caption{Our method is robust despite extreme initial conditions: a randomly initialized hippo
returns back to its rest pose.}
\label{fig:hippo_recover}
\end{figure}

{\bf Robustness.} We demonstrate that our proposed extensions to more general materials and the
L-BFGS solver upgrade do not compromise simulation robustness. In \reffig{hippo_recover}, we show
an elastic hippo which recovers from an extreme (randomized) deformation with many inverted
elements. Specifically, the hippo model uses L-BFGS with $m=5$ and corotated linear elasticity with
$\mu = 20$ and $\lambda = 100$ (note that Projective Dynamics supports only corotated materials
with $\lambda = 0$).

\vspace{-0.18cm}
\section{Limitations and Future Work}
\vspace{-0.18cm}

Our method is currently limited only to hyperelastic materials satisfying the Valanis-Landel
assumption. Even though this assumption covers many practical models, including the recently
proposed spline-based materials \cite{xu2015nonlinear}, it would be interesting to study the further
generalization of our method. Perhaps even more interesting would be to remove the assumption of
hyperelasticity. Can we develop fast algorithms for simulating non-hyperelastic materials,
including the effects such as relaxation, creep, and hysteresis \cite{bargteil2007finite}?
Inspired by the recent work of
Wang~\shortcite{wang2015chebyshev}, we would like to explore GPU implementations of physics-based
simulations. Our current method is derived from the Implicit Euler time integration method and
therefore inherits its artificial damping drawbacks. We experimented with Implicit Midpoint -- a
symplectic integrator which does not suffer from this problem. However, we found that Implicit
Midpoint is much less stable. In the future we would like to explore fast numerical solvers for
symplectic yet stable integration methods. Finally, we plan to investigate specific physics-based
applications which require \textit{both} high accuracy and speed, such as interactive surgery
simulation.


%
%




\vspace{-0.18cm}
\section{Conclusions}
\vspace{-0.2cm}

We have presented a method for fast physics-based simulation of a large class of hyperelastic
materials. The key to our approach is the insight that Projective Dynamics
\cite{Bouaziz2014Projective} can be re-formulated as a quasi-Newton method. Aided with line search,
we obtain a robust simulator supporting many practical material models. Our quasi-Newton
formulation also allows us to further accelerate convergence by combining our method with L-BFGS.
Even though L-BFGS is sensitive to initial Hessian approximation, our method suggests a
particularly effective Hessian initialization which yields fast convergence. Most of our
experiments use ten iterations of our method which is typically more accurate than one iteration of
Newton's method, while being about ten times faster and easier to implement. Traditionally,
real-time physics is considered to be approximate but fast, while off-line physics is accurate but
slow. We hope that our method will help to blur the boundaries between real-time and off-line
physics-based animation.

\bibliographystyle{acmsiggraph}
\bibliography{genPD}

\vspace{-1mm}
\section*{Appendix}
\vspace{-1mm}

In this Appendix we compute the gradient $\nabla_\x E_i(\x)$, where $E_i(\x)$ is defined according
to \refequ{PDenergy}. Dropping the subscript $i$ for clarity, we define the projection:
\begin{equation}\label{equ:minProj}
\p(\x) = \argmin_{\z \in \cM} \| \G \x - \z \|^2
\end{equation}
where $\G$ represents a discrete differential operator and $\cM$ is a constraint manifold. We need
to compute the differential:
\begin{eqnarray}
\!\!\!\!\!\!\!\!\! \frac{1}{2} \delta \| \G \x - \p(\x) \|^2 &=& (\G \x - \p(\x))^\tr (\G \delta \x - \delta \p(\x)) \\
\!\!\!\!\!\!\!\!\! &=& (\G \x - \p(\x))^\tr \G \delta \x
\end{eqnarray}
because the second term $(\G \x - \p(\x))^\tr \delta \p(\x)$ vanishes. This is due to the fact that
$\delta \p(\x) \in T_{\p(\x)} \cM$, where $T_{\p(\x)} \cM$ denotes the tangent space at point
$\p(\x) \in \cM$. The vector $\G \x - \p(\x)$ must be orthogonal to $T_{\p(\x)} \cM$, otherwise
$\p(\x)$ could not be the minimizer according to \refequ{minProj}.

\end{document}